\documentclass[%
 reprint,
amsmath,amssymb,
aps,
]{revtex4-2}

\usepackage{graphicx}
\usepackage{dcolumn}
\usepackage{float}
\usepackage{comment}
\usepackage{bm}
\usepackage{color, soul}
\usepackage{physics}
\usepackage{chngcntr}
\usepackage[caption=false]{subfig}
\newcommand{\mar}[1]{{\color{red}#1}}

\begin{document}

\preprint{APS/123-QED}

\title{Leveraging collective effects for thermometry in waveguide quantum electrodynamics}

\author{Aleksei Sharafiev$^{1}$}
\author{Mathieu Juan$^{2}$}
\author{Marco Cattaneo$^{3}$}%
\author{Gerhard Kirchmair$^{1,4}$}%
 \email{gerhard.kirchmair@uibk.ac.at}
\affiliation{$^{1}$Institute for Quantum Optics and Quantum Information, Austrian Academy of Sciences, 6020 Innsbruck, Austria}

\affiliation{$^{2}$Institut Quantique and D\'{e}partement de Physique,
Universit\'{e} de Sherbrooke, Sherbrooke, Qu\'{e}bec, J1K 2R1, Canada}
\affiliation{$^{3}$QTF Centre of Excellence, Department of Physics,
University of Helsinki, P.O. Box 43, FI-00014 Helsinki, Finland}
\affiliation{$^{4}$Institute for Experimental Physics, University of Innsbruck, 6020 Innsbruck, Austria}

\date{\today}

\begin{abstract}
We report a proof-of-principle experiment for a new method of temperature measurements in waveguide quantum electrodynamics (wQED) experiments, allowing one to  differentiate between global and local baths. The method takes advantage of collective states of two transmon qubits located in the center of a waveguide. The Hilbert space of such a system forms two separate subspaces (bright and dark) which are coupled differently to external noise sources. Measuring transmission through the waveguide allows one to extract separately the temperatures of the baths responsible for global and local excitations in the system. Such a system would allow for building a new type of primary temperature sensor capable of distinguishing between local and global baths.
\end{abstract}

\maketitle

Characterizing and controlling the coupling between quantum systems and environmental degrees of freedom is one of the central problems in modern quantum systems engineering \cite{murch_cavity-assisted_2012,soare_experimental_2014,aron_steady-state_2014,kimchi-schwartz_stabilizing_2016,shabani_artificial_2016,campbell_global_2017,harrington_bath_2019,latune_collective_2020,cattaneo_bath-induced_2021,kitzman_phononic_2023}. In some cases, this coupling might be useful as in quantum sensing \cite{degen_quantum_2017}, quantum thermometry \cite{de_pasquale_quantum_2018} or quantum engines \cite{cangemi_quantum_2023}, while in others it is harmful e.g. for quantum computation protocols, leading to an additional dephasing as well as excessive qubit population. The coupling of one quantum system to multiple baths
attracted significant attention during the last years both on theoretical and experimental sides, including the field of superconducting quantum circuits \cite{Karimi2017,hu_steady-state_2018,hofer_markovian_2017,cattaneo_local_2019,cattaneo_engineering_2021,ronzani_tunable_2018,von_lupke_two-qubit_2019,senior_heat_2020}.

Thermalization of a many body system coupled to various baths is especially relevant in the context of superconducting devices whose effective temperatures are known to plateau at 30-50~mK, leading in particular to excess qubit population \cite{corcoles_protecting_2011,jin_thermal_2015,kulikov_measuring_2020}. Various possible reasons for this behaviour have been revealed, including poorly thermalized attenuators in the input lines of the cryostat \cite{yeh_microwave_2017,krinner_engineering_2019} as well as nonequilibrium quasiparticles, perhaps produced by stray infrared radiation \cite{barends_minimizing_2011} or cosmic rays \cite{henriques_phonon_2019, mcewen_resolving_2022}. Continuing research in this area during the last few years resulted in several proposed and realized protocols for qubit temperature measurements \cite{jin_thermal_2015,krantz_single-shot_2016,kulikov_measuring_2020} and new techniques for its reduction starting to appear \cite{lucas_quantum_2023}.     

To understand the thermal baths a superconducting circuit is coupled to one has to perform thermometry on the circuits. Several protocols have been recently proposed \cite{karimi_noninvasive_2018,Wang2018,wheeler_sub-kelvin_2020,Blanchet2022} including measuring the  temperature of the environmental photonic bath through qubit population \cite{scigliuzzo_primary_2020}. 
The field of thermometry recently expanded to the quantum regime, where entanglement can be used to overcome fundamental limitations of classical thermometry \cite{de_pasquale_quantum_2018}. For instance, exploiting bath-induced correlations may enhance thermometry precision \cite{planella_bath-induced_2022}.
\begin{figure}
    \centering
    \includegraphics[width=1\linewidth]{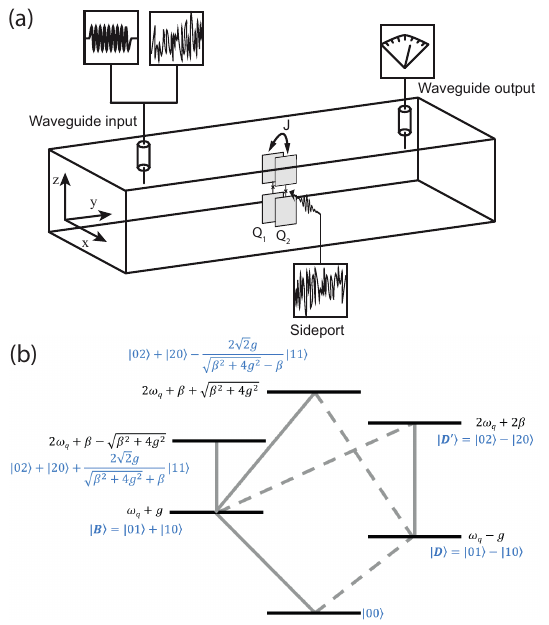}
    \caption{(a) Experimental setup: a rectangular waveguide (global bath) with two transmon qubits inside; a pin is insterted on one side of the waveguide and connected to a sideport (local bath). (b) Simplified level scheme of the two transmons with the corresponding wavefunctions (normalization omitted). Solid arrows correspond to allowed transitions in the global regime, while dashed lines to prohibited ones. Notations: $\beta<0$ - anharmonicity of the transmons, $g>0$ - capacitive coupling between the transmons.}
    \label{fig:intro}
\end{figure}

This work introduces a new method for the individual extraction of temperature associated with two noise sources affecting a system of two superconducting transmons, and demonstrates it through a proof-of-principle experiment. Leveraging the inherent symmetry of the system, our approach employs multi-probe spectroscopy to distinguish between a collective, global bath and a more localized noise source. In this context, we will employ the term \textit{global} to denote a thermal reservoir symmetrically coupled to both qubits, whereas \textit{local} refers to a noise source that is more strongly coupled to one qubit, thereby inducing an asymmetry in the dissipation channels. The latter scenario encompasses purely localized baths extensively examined in theoretical studies \cite{Levy2014,cattaneo_local_2019,Trushechkin2016,hofer_markovian_2017,Gonzalez2017}, wherein noise sources exclusively affect a single qubit.

Our method introduces a significant contribution to the toolbox of existing cryogenic thermometry protocols, offering a novel capability of distinguishing between local and global noise source. Therefore, our approach holds the potential to emerge as an indispensable tool for detecting and discerning the different noise sources influencing the behavior of superconducting qubits. Such discernment is paramount for the strategic design of possible countermeasures and better understand the complex thermalization of extended circuits.

Our experimental setup is shown in the Fig.~\ref{fig:intro}(a) where two capacitively coupled flux tunable transmon qubits ($\text{Q}_1$ and $\text{Q}_2$) are located in the center of a rectangular cross-section waveguide (the setup is similar to what was used in \cite{sharafiev2021visualizingemission,zanner_coherent_2022}). The dimensions of the waveguide determine the frequency range over which the qubits can be tuned while remaining in the single mode regime of the waveguide.
For the experiment, the system was equipped with two electromagnets thermally anchored to a base plate of a dilution refrigerator (not shown), allowing for an independent control of the qubits frequencies. The transmons were fabricated with a standard  Al/AlO$_x$/Al bridge-free technique \cite{lecocq_junction_2011} on 330$\mu$m-thick sapphire substrate. 

Owing to the symmetry of the system, we separate the two-qubit Hilbert space into a \textit{dark} or \textit{subradiant} subspace, which is effectively decoupled from the noise coming through the waveguide, and a \textit{bright} or \textit{superradiant} subspace \cite{cattaneo_bath-induced_2021,Karimi2017,liao_quantum_2011} (see Fig.~\ref{fig:intro}(b)). In addition, a pin mounted on the side all of the waveguide weakly couples to both qubits. This line is used to engineer a controlled thermal bath acting on the qubits, whose effective temperature can be varied in a broad range (roughly from 100~mK to 60~K with the noise source we used). Crucially, owing to the asymmetry of the evanescent field of the pin, this bath couples to both dark and bright subspaces. 

In the following, the bath associated with the pin line will be referred to as \textit{local} in contrast to the bath associated with the fundamental mode of the waveguide, which we will call \textit{global}. An alternative attempt to engineer a local bath through a resistive heater can be found in Appendix~\ref{sec:heater_section}. The system was designed in a way to have unequal couplings to different baths: $\gamma_{loc1}<\gamma_{loc2}\ll\gamma_{glob}$. A more strongly coupled global bath corresponds to a typical experimental situation in waveguide QED when local baths can be associated with internal losses. 
Having the qubit pair coupled strongly to the global bath of a continuous mode environment makes the system more sensitive to its temperature, in turn leading to a higher residual temperature for the qubits. For this proof-of-principle experiment this situation is acceptable as this residual temperature, associated with the microwave photons in the input line, can be easily included in the model. 

The system illustrated in Fig.~\ref{fig:intro}(a) has 3 ports: input and output of the waveguide, connected correspondingly to standard 50~$\Omega$ input and output lines, and the side pin line, connected to a second standard 50~$\Omega$ input line. All input lines in our setup are equipped with 60~dB attenuation, while the output line has a 40~dBm high-electron-mobility transistor (HEMT) amplifier along with 2 circulators between the HEMT and the waveguide, to prevent possible noise propagation from the HEMT to the qubits. To couple our system to a bath (local or global) we apply controlled white noise to the corresponding port, produced by amplifying the Johnson noise of a 50~$\Omega$ load with a room temperature amplifier chain combined with a digital attenuator (see Appendix~\ref{sec:details_of_exp} for details). By changing the settings of the attenuator we effectively regulate the noise temperature. 

As we can send noise separately to the different input ports, we are able to independently study the global and local baths. The effect of each bath is obtained by measuring the transmission through the waveguide which depends on the steady state of the qubits~\cite{lalumiere_input-output_2013}, and thus their temperatures. Due to the presence of a dark subspace with respect to the global case, the transmission coefficient in this scenario will be crucially different from the local bath case where the dark state is broken. Through a suitable fit of these coefficients we can then extract the temperatures of the baths independently. 

For a theoretical description of our model (see details in Appendix~\ref{sec:theory_section}), we will model the effect of the baths on the two-qubit system through a Bloch-Redfield master equation in a partial secular approximation \cite{BreuerPetruccione,cattaneo_local_2019}, whose Markovianity is justified by the approximately flat spectral density of the baths (see Appendix~\ref{sec:spectrum_consideration} for some related considerations) and by the weak coupling between qubits and reservoirs, combined with input-output formalism \cite{lalumiere_input-output_2013}. 
\begin{figure*}
    \centering
    \includegraphics[width=1\linewidth]{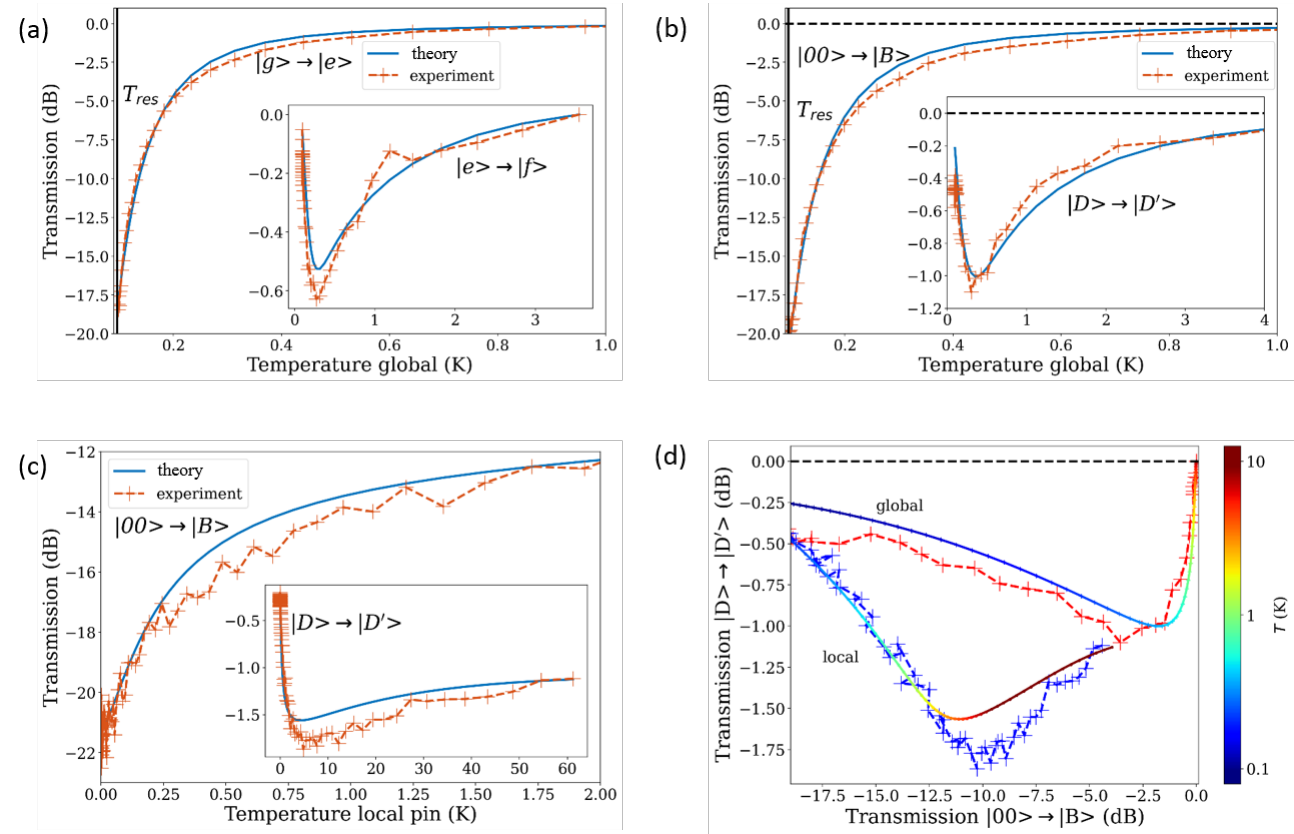}
    \caption{\label{fig:main_results}   Transmission measurements (dashed red) and corresponding fitting curves (solid blue) for: (a) single qubit, only global bath applied, (b) two qubits in resonance, only global bath applied, and (c) two qubits in resonance, only local bath applied. The main plots correspond to the transitions $\ket{g}\rightarrow\ket{e}$ or $|00>\rightarrow\ket{B}$ while the insets to $\ket{e}\rightarrow\ket{f}$ and $\ket{D}\rightarrow\ket{D'}$. (d) Bright-dark diagram obtained from (b) and (c). Dashed red (blue): transmission measurements with a global (local) bath. The color of the solid lines (theoretical predictions) reflects the temperature of the corresponding bath.}
\end{figure*}
We can understand the behavior of the transmission coefficients in the global and local scenarios by looking at the simplified level scheme in Fig.~\ref{fig:intro}(b) (a more detailed scheme can be found in Appendix~\ref{sec:spectrum_consideration}): under the action of a perfectly symmetric global bath, the levels $\ket{D}$ and $\ket{D'}$ are decoupled from the other levels as the transitions are prohibited by symmetry. These levels therefore form a dark subspace and will not be populated by a global bath coupled only to the bright subspace, regardless of the temperature of the bath. On the other hand, a bath which is coupled only to one qubit (or asymmetrically to both qubits) breaks the symmetry and therefore can populate the $\ket{D}$ state. 

Note that the $\ket{D}\rightarrow\ket{D'}$ transition is allowed even under the action of a symmetric global bath, and therefore can be probed through transmission measurements. The visibility of this transistion depends directly on the steady state population of the state $\ket{D}$. As a result, the transmission through the waveguide for frequencies corresponding to the $\ket{00}\rightarrow\ket{B}$ (in the bright subspace) and $\ket{D}\rightarrow\ket{D'}$ (in the dark subspace) transitions can be used to extract independently the temperatures of both global and local baths. Remarkably, the argument still holds in non-ideal scenarios. In our experiment, the anharmonicities of the qubits are non identical \cite{koch_charge-insensitive_2007}, therefore the coupling to the global bath is not perfectly symmetric. However, we will show that the independent temperature estimation is possible even for such a non-ideal real-life setup.  

First, we calibrate the power reaching the qubits from the 2 input ports, using a simple Autler-Towns experiment (see Appendix~\ref{sec:pow_cal}). For this, we apply a coherent tone to the corresponding input in resonance with one of the qubits while the other qubit is detuned below the cutoff of the waveguide. As a result, the qubit frequency splits and the width of the splitting is proportional to the Rabi frequency. Then, we fit the effects of the noise source power on one of the qubits ($\text{Q}_2$) to calibrate the effective temperature of the bath. The model has 2 fitting parameters: residual temperature $T_{res}$ and conversion coefficient $\alpha$ between the power of the source and added noise temperature, i.e. $T=T_{res}+\alpha \cdot P$, where $P$ is the power we send into the fridge.
We estimate these values by getting the best possible fit of our theoretical model to the experimental transmission coefficients as a function of temperature of the global bath for the main transition $\ket{g}\rightarrow\ket{e}$  (see Appendix~\ref{sec:theory_section}) for a bath with temperature $T$. The results of the experiment and their best theoretical fit are shown in Fig.~\ref{fig:main_results}(a). The extracted value of $T_{res}$ is 95~mK with a base temperature of the fridge of 14~mK. We associate this relatively high value with the influence of a residual global bath - photons coming from the input port of the waveguide. 
This result is expected in waveguide QED experiments, where the qubits are placed in a continuous 50~$\Omega$ environment.

Then, we compute the transmission coefficients of different transitions for the single and two qubit system as a function of temperature using our theoretical model with the fit values for $T_{res}$ and $\alpha$, and compare them with the experimental results. The good agreement between the experimental and theoretical lines shows that our theoretical model is a good description of the experiment, as can be seen for instance in the inset in Fig.~\ref{fig:main_results}(a) for the transmission coefficients of the $\ket{e}\rightarrow\ket{f}$ transition or Fig.~\ref{fig:main_results}(b) for the $\ket{00}\rightarrow\ket{B}$ transition.


Next, we set both qubits in resonance at 7.8~GHz  and vary the noise temperature of both global and local baths. In this case we are interested in 2 transitions: $\ket{00}\rightarrow\ket{B}$ and $\ket{D}\rightarrow\ket{D'}$, see Fig.~\ref{fig:intro}(a). We start by varying the temperature of the global bath and monitoring the transmission at the $f_{00B}$ and $f_{DD'}$ frequencies with a weak probe. The result of the experiment and the corresponding theoretical lines are shown in Fig.~\ref{fig:main_results}(b). 

In a perfect device, the  global bath temperature does not affect the transmission through the waveguide at the $\ket{D}\rightarrow\ket{D'}$ frequency $f_{DD'}$, as the global bath is decoupled from the dark subspace. Nevertheless a dependence on temperature appears both in the experiment and the theoretical fit, as demonstrated in the inset of Fig.~\ref{fig:main_results}(b). However, the change in intensity is much lower than for the $\ket{00}\rightarrow\ket{B}$ transition. There are two reasons for this dependence. First, the spectrum of allowed transitions in the bright subspace is richer than the one shown in the Fig.~\ref{fig:intro}(b), and it includes transitions which are close to $f_{DD'}$ in higher excitation manifolds, see Appendix~\ref{sec:spectrum_consideration}. Second, no  physical system displays perfect symmetry. In our experiment the symmetry is broken by the imprecise location of the qubit pair inside the waveguide and by slightly different anharmonicities of the transmons $\beta$. In our setup, the qubits have anharmonicities $\beta_1\sim-225$~MHz and $\beta_1\sim-232$~MHz respectively for $\text{Q}_1$ and $\text{Q}_2$, i.e. their difference is of the order of $\sim2.6\%$. Spectroscopic lines appearing in the system with up to 3 excitations and the difference in $\beta$ between the qubits have been taken into account in the model. As one can see from the figure, the observed transmission at $f_{DD'}$ is well described by the curve obtained with the theoretical model.

Finally we use the side pin to apply a local bath, affecting the qubits asymmetrically. 
Performing an Autler-Towns experiment, but applying the pump tone from the side pin, we were able to extract the ratio of the couplings between the qubits and the side pin, obtaining $\gamma_{loc2}/\gamma_{loc1}=1.78$. This ratio was used in the model, while the coupling itself was kept as a fitting parameter. To extract the coupling we again fit only single-qubit data, which in this case is fitted simultaneously for both qubits. For the details on this procedure see Appendix~\ref{sec:local_fit_section}. After extracting the coupling, we plot the experimental and theoretical results for the two qubits and local bath experiment in  Fig.~\ref{fig:main_results}(c). As before, the theoretical lines are obtained with the values of $T_{res}$ and $\alpha$ fitted in single-qubit experiments.

Comparing Fig.~\ref{fig:main_results}(b) with Fig.~\ref{fig:main_results}(c) one can see that indeed the effect on the $\ket{D}\rightarrow\ket{D'}$ transition is more noticeable in the local bath case: the transmission drop is deeper and it occurs when the $\ket{00}\rightarrow\ket{B}$ transmission is far from reaching saturation. This can be better illustrated with what we call \textit{bright-dark diagram}, see Fig.~\ref{fig:main_results}(d), where the axes  correspond to the transmission values at the frequencies $f_{00B}$ and $f_{DD'}$. Experimental frequency sweeps in Appendix~\ref{sec:heater_section} further illustrate the difference between the effects of global and local baths on the transmission.

From this diagram one can see that local and global bath saturate the  transitions $\ket{D}\rightarrow\ket{D'}$ and $\ket{00}\rightarrow\ket{B}$ very differently. Thus, a combined measurement allows us to assess the influence of a local bath on a temperature measurement using the $\ket{00}\rightarrow\ket{B}$ as a calibrated sensor. In our proof of principle experiment, where both baths have been calibrated, we can actually determine the temperature of both uniquely by measuring the depth of both spectroscopy peaks. This works for a temperature range of 50~mK to about 500~mK for the global and 100~mK to about 2~K for the local baths. In a real world scenario one would not be able to determine the temperature of the local bath as the coupling would not be known. In this case our method would still allow to extract its influence on the measurement of the global temperature (where the system still acts as a calibrated sensor) and correct for it.

\begin{figure}[h]
\includegraphics[width=1\linewidth]{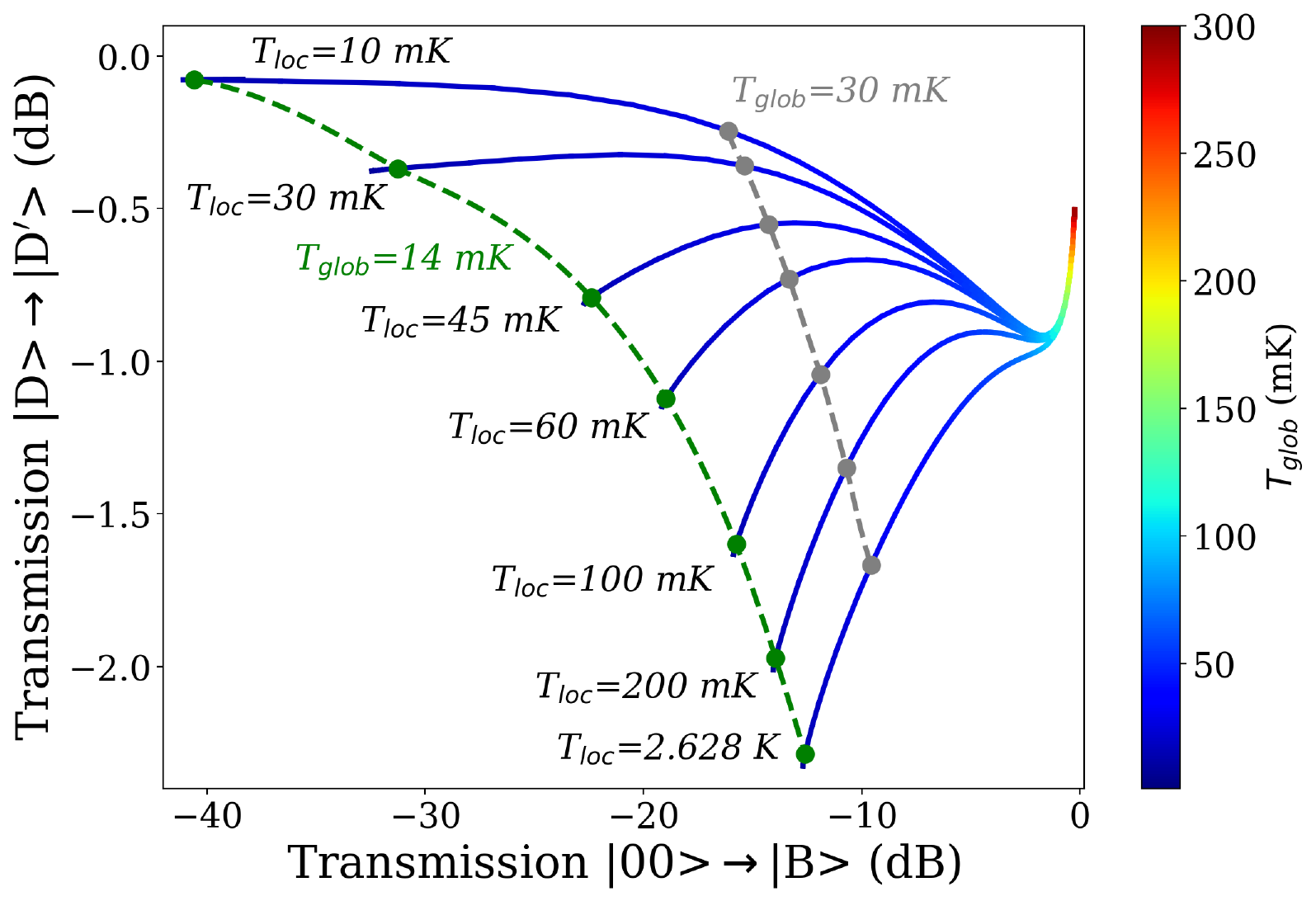}
\caption{\label{fig:last} Theoretical bright-dark diagrams for an experiment with low frequency (2~GHz) transmons. All lines correspond to $T_{glob}$ change from 10 to 300~mK. The shape of the curves allows us to reveal a weakly coupled local bath even when the residual temperature of the qubits is defined by the global bath (see the main text).} 
\end{figure}

To illustrate the measurement process further we have run simulations of a system based on our experiment that is suitable to detect the desired low input mode temperatures ($\approx 20mK$) of a circuit quantum electrodynamics system. The results of this simulation can be seen in Fig.~\ref{fig:last}. Here we change the global bath temperature from 10 to 300~mK for a pair of 2~GHz transmons inside a waveguide. We assume a local bath coupling of 1~kHz with other parameters corresponding to our current experiment. Again, a combined measurement of both transitions allows to determine the temperature of the global bath and extract the influence of the local bath. 

From the level diagram (see Fig.~\ref{fig:intro}(b)), one can see that the local bath has a very strong influence on the temperature measurement using the transmission on the $\ket{00}\rightarrow\ket{B}$ leading to a wrong inference of the global bath temperature when only this transition is used. This is the same situation for thermometry methods based on a single qubit: local and global bath cannot be differentiated making it impossible to definitively identify the origin of excess temperature in the system.  From the measurement of the $\ket{D}\rightarrow\ket{D'}$ one can determine the residual excitation of the qubits due to a local baths and correct for it in the global bath measurement. In other words for a well calibrated two-qubit sensor with a verified model, each ``line" in the dark-bright diagram can be uniquely associated with a combined measurement of the  $\ket{D}\rightarrow\ket{D'}$ and $\ket{00}\rightarrow\ket{B}$ lines. Two dashed lines in Fig.~\ref{fig:last} corresponding to global bath temperature 14 and 30~mK serve as guide to the eye and correspond to typical base plate temperature of a dilution refrigerator and a lowest residual transmon frequency typically observed experimentally. Some more details are provided in Appendix~\ref{sec:model_results}.

In conclusion, we have demonstrated a proof-of-principle setup that allows to distinguish between two types of baths coupled to a quantum sensor by measuring two transitions of the coupled system. We achieve this by exploiting system symmetries combined with a good calibration of the system and a working theory model. Developing this proof-of-principle experiment into a useful thermometer that would accurately measure the input mode temperature of a system and be able to correct for undesired local bath couplings would require switching to a planar CPW architecture as manufacturing tolerances leading to asymmetries and impedance matching could be improved. We believe the approach presented in this paper opens a promising avenue for a new type of primary temperature sensor capable of distinguishing between local and global baths for temperatures as low as $\sim$20-30~mK enabling more accurate temperature measurements.

\textit{Acknowledgments}.--The authors would like to thank Roberta Zambrini for useful comments on the theoretical model employed in this work. MC would like to thank the members of the PICO group at Aalto University for interesting discussions and suggestions. AS and GK acknowledge funding by the European Research Council (ERC) under the European Unions Horizon 2020 research and innovation program (714235). This research was funded in part by the Austrian Science Fund (FWF) DOI 10.55776/F71. MC acknowledges funding from the Research Council of Finland Centre of Excellence program grant 336810 and from COQUSY project PID2022-140506NB-C21 funded by MCIN/AEI/10.13039/501100011033. MLJ acknowledges funding by the Canada First Research Excellence Fund.

\bibliography{thermometry_paper_refs}

\newpage

\appendix

\newpage    
\begin{widetext}
 \counterwithin{figure}{section}
 
\section{Adding a heater}
\label{sec:heater_section}
A naive way to create a local bath would be adding a physical heater to the chip. With the aim to compare this approach with our model, we originally designed the setup with the qubits located on 2 different and rather long - 25~mm - chips. The chips protrude outside of the waveguide and the heater (SMD resistor) was varnish-glued on the lower end of one of them, as shown in Fig.~\ref{fig:wg_and_heater}\subref{fig:wg_cross_section}. The chips were only connected mechanically on the upper end with a copper clamp. The resistor was wired to a current source to regulate its temperature. As the resistor was located around 1~cm away from the SQUID loops of the qubits, we could be confident that the small heating current (up to 100$\mu$A) would not affect qubit frequencies. It is worth to note that this experiment is different from an approach used in some other works, e.g. \cite{ronzani_tunable_2018}, where resistor was used just as a capacitively coupled bath (similarly to our side pin experiment). It is also different from heating the base plate of the refrigerator, used e.g. in \cite{scigliuzzo_primary_2020}, as it effectively leads to heating the attenuators in the input lines and therefore would change the temperature of the global bath.

The experiment with the heater revealed that our model was not capable of predicting the systems behaviour mainly because of significant change of the junctions critical currents. In turn, it caused a qubit frequency change which complicated extracting the effect of the qubits populations on the transmission and would require a more complex model. The Fig.~\ref{fig:wg_and_heater}\subref{fig:fr_shift} shows the change of the lowest transmission point in a single qubit experiments for a local bath delivered by the side pin, local bath as a heater and prediction of our model. The discrepancy in the low attenuation (high temperature) area between the model and the experiment in our case was around 4~MHz and probably associated with the fact that only 4 levels of the transmon were considered in the theory. The discrepancy for the heater bath due to critical current change was around 40~MHz, i.e. larger than the qubits coupling to the waveguide. Moreover, thermal isolation of the qubits turned out to be worse than we expected, and heating effect could be simultaneously seen on both qubits. In this case the fitting procedure with a fixed probe frequency used in the main text could not be reasonably applied.

Nevertheless, to visualize the difference between three baths (global, local-pin and local-heater) we conducted a frequency sweep of qubit $\text{Q}_1$ while keeping qubit $\text{Q}_2$ fixed, see Fig.~\ref{fig:flux_sweeps}. In all three cases the bath temperature has been chosen to maximize the effect on the $\ket{D}\rightarrow\ket{D^{\prime}}$ transition, while the color scale is the same. The difference between the effects of global and local-pin baths can be clearly seen from Fig.~\ref{fig:flux_sweeps}\subref{fig:global_sweep_3D} and Fig.~\ref{fig:flux_sweeps}\subref{fig:pin_sweep_3D}.

\begin{figure}[h]
\subfloat[\label{fig:wg_cross_section}]{%
\includegraphics[scale=1.2]{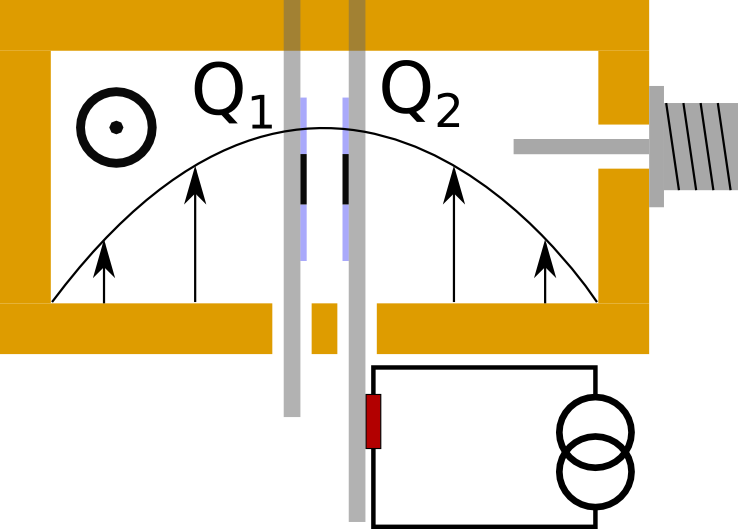}%
}\hfill
\subfloat[\label{fig:fr_shift}]{%
\includegraphics[scale=0.33]{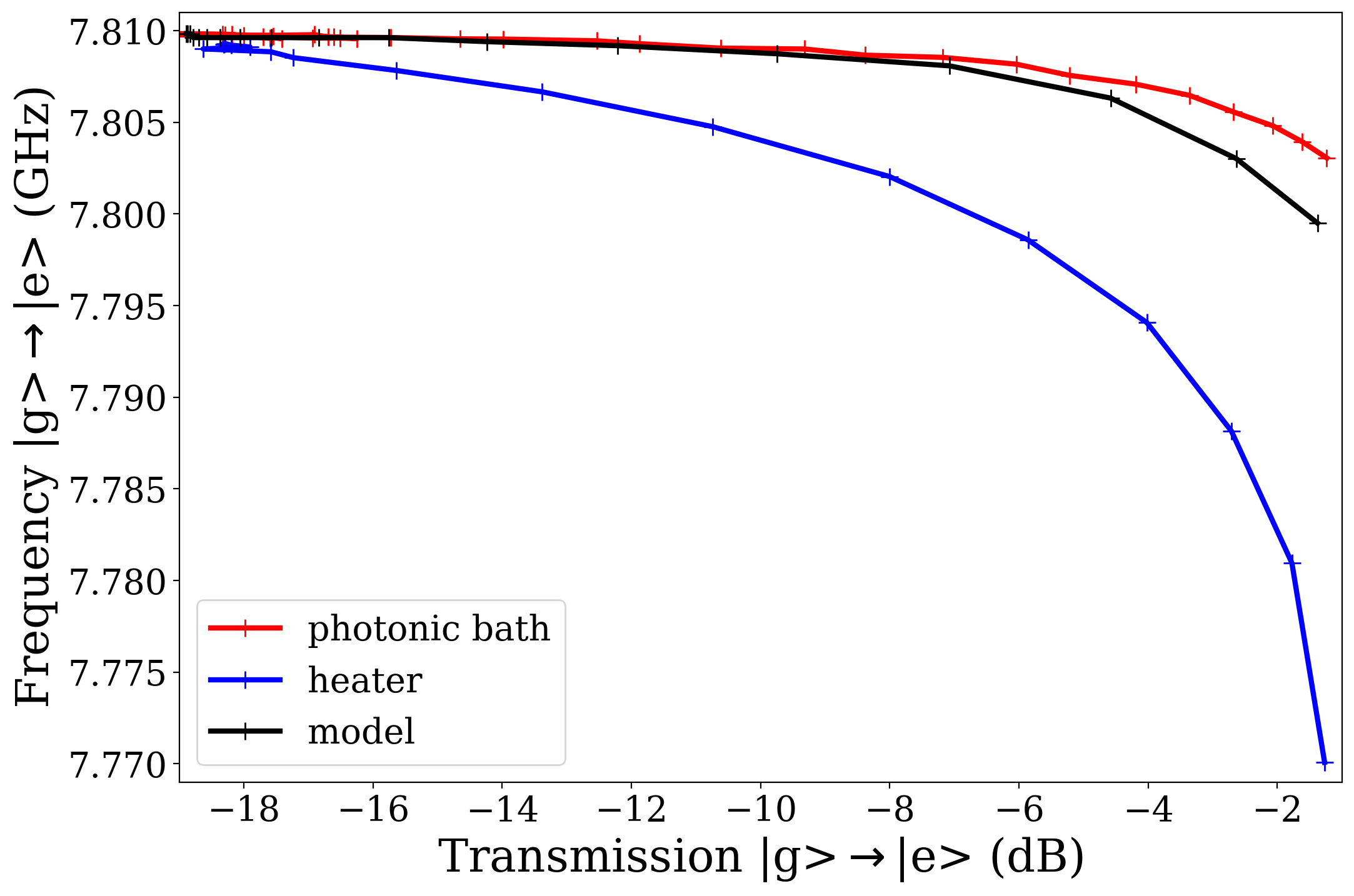}%
}\hfill
\caption{\label{fig:wg_and_heater} (a) Cross-section of the waveguide with 2 qubits ($\text{Q}_1$ and $\text{Q}_2$) inside and a resistive heater attached to one of them.  (b) The effect of local bath heating on the lowest transmission frequency corresponding to $\ket{g}\rightarrow\ket{e}$ transition. Horizontal axis is the value of the transmission.}
\end{figure}

\begin{figure}[H]
\centering
\subfloat[\label{fig:global_sweep_3D}]{%
\includegraphics[scale=0.25]{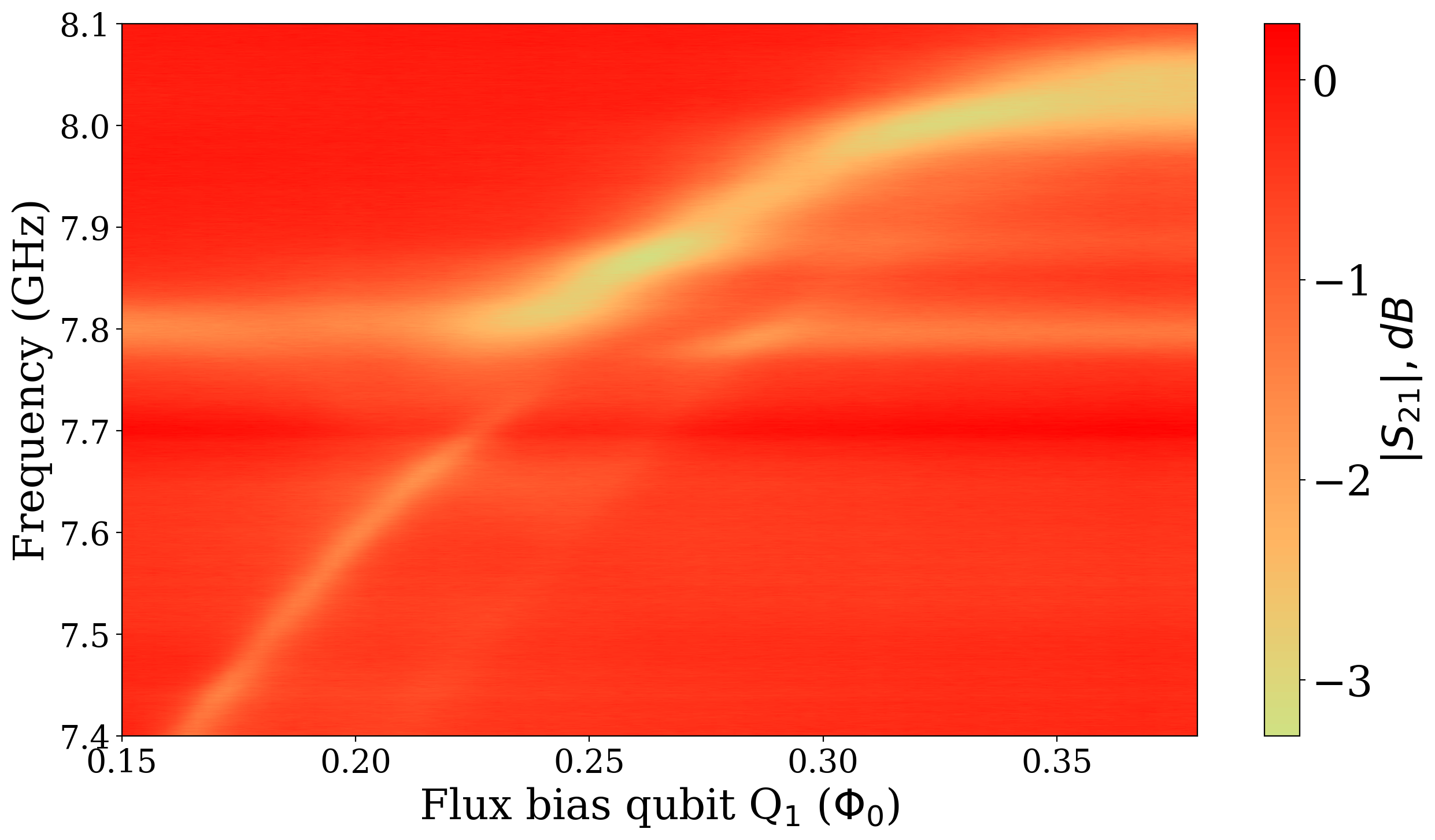}%
}
\subfloat[\label{fig:pin_sweep_3D}]{%
\includegraphics[scale=0.25]{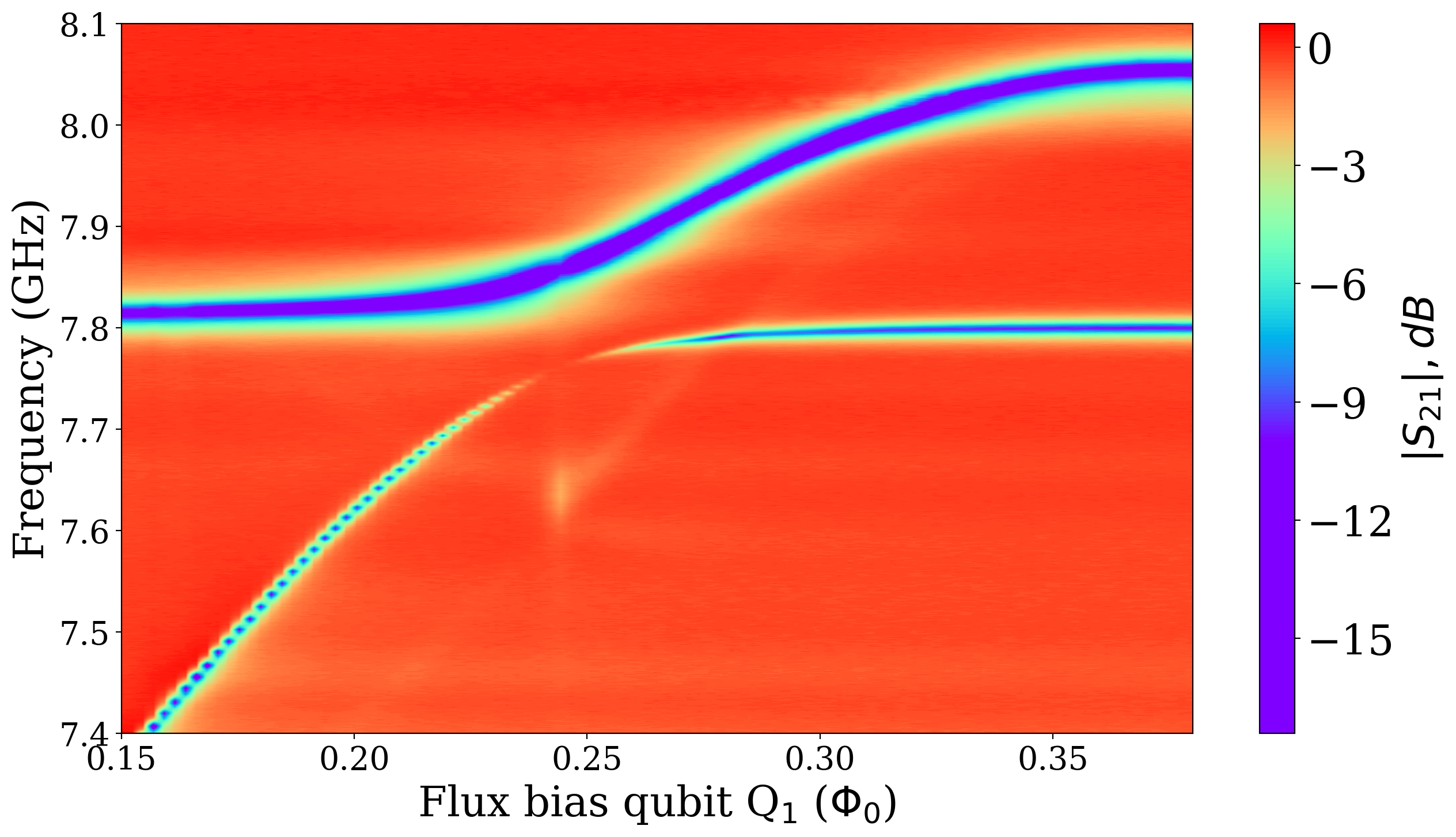}%
}\hfill
\subfloat[\label{fig:heater_sweep_3D}]{%
\includegraphics[scale=0.25]{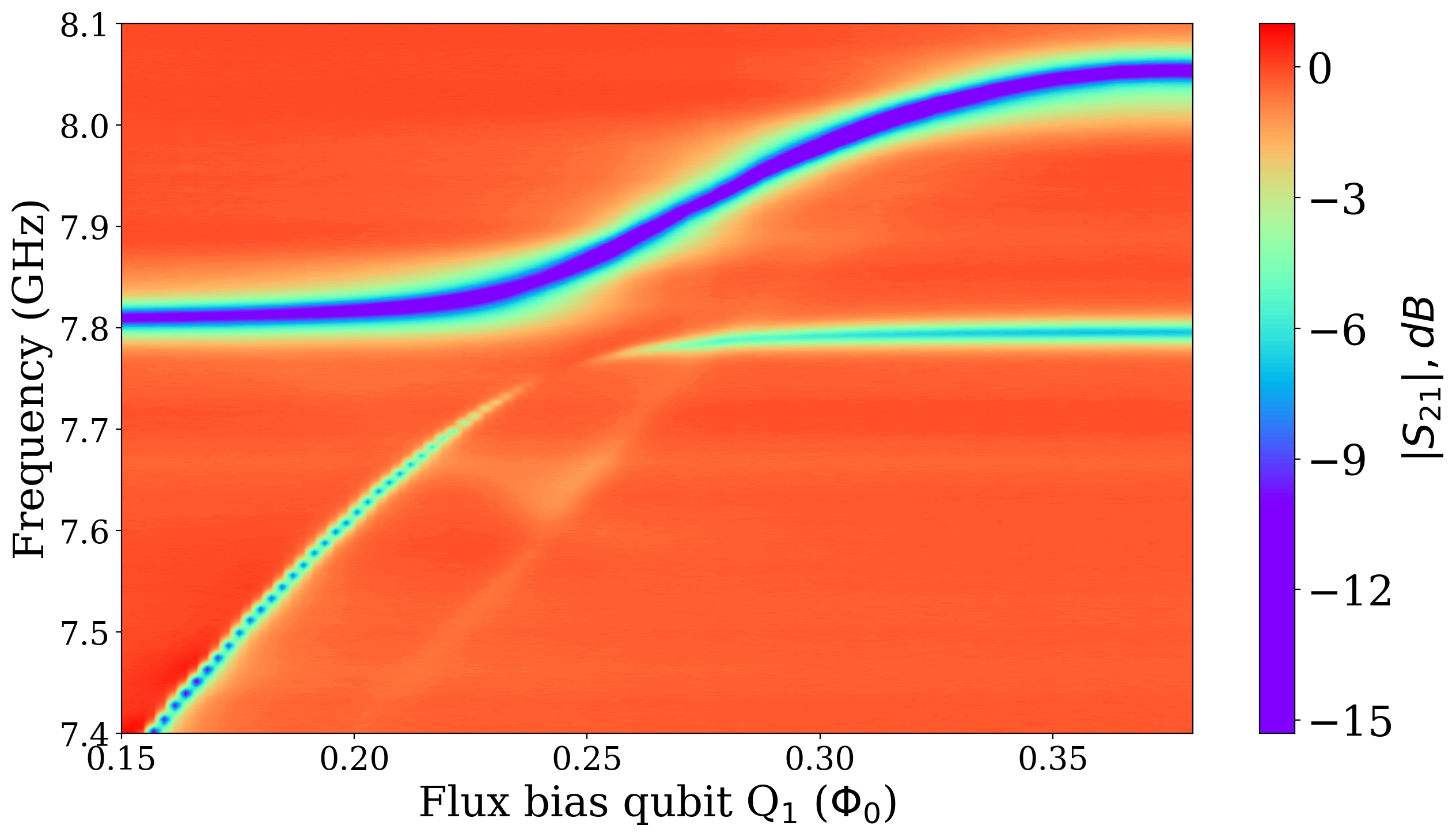}%
}
\caption{\label{fig:flux_sweeps} Transmission measurements with a weak signal through the waveguide while sweeping the frequency of the transmon $\text{Q}_1$, keeping transmon $\text{Q}_2$ still for 3 different situations: (a) only global bath is on; (b) only local bath generated by a side pin is on; (c) only local bath generated by a heater is on.}
\end{figure}

\section{Details of the experimental setup}
\label{sec:details_of_exp}
The experimental setup schematic is shown in the Fig.~\ref{fig:heater_pic}\subref{fig:setup_scheme}. We use 2 input and 1 output lines inside a dilution refrigerator. The input lines have 60~dB attenuation and the output line is equipped with a low-noise HEMT 
amplifier and a room temperature amplifier, providing +40~dB amplification each. Noise was applied from a room temperature amplifier terminated with 50~$\Omega$ on the input, through a variable attenuator, effectively regulating the noise temperature. A probe tone was applied and measured with a standard Vector Network Analyzer (VNA). Current sources were used to independently tune qubits frequencies with magnetic field and apply current through the heating resistor.

Both qubits inside the waveguide (WG) were flux tunable approximately between 4 and 8~GHz, and were fabricated with a standard Al/AlO$_{x}$/Al bridge-free technique \cite{lecocq_junction_2011} on 330$\mu$m-thick sapphire substrate.

The spectrum of the applied noise was measured separately with a spectrum analyzer and shown with a blue line in the Fig.~\ref{fig:heater_pic}\subref{fig:att_test_and_noise}. The noise spectral density was normalized to its value at the $\ket{00}\rightarrow\ket{B}$ transition frequency.

As we learned, many variable attenuator models suffer from substantial ripples especially at large attenuation values. To make sure the ripples are not significant in our case, we checked how flat the attenuation spectrum is. Red lines in the Fig.~\ref{fig:heater_pic}\subref{fig:att_test_and_noise} correspond to the accuracy of the attenuation settings, i.e. difference between set and measured values. One can see that the noise spectral density as well as the imperfections of the attenuator do not cause major deviations of the noise from being relatively white at least at 2 frequencies of interest corresponding to the $\ket{00}\rightarrow\ket{B}$ and $\ket{D}\rightarrow\ket{D^{\prime}}$ transitions.      
\begin{figure}[H]
\subfloat[\label{fig:setup_scheme}]{%
\includegraphics[scale=0.6]{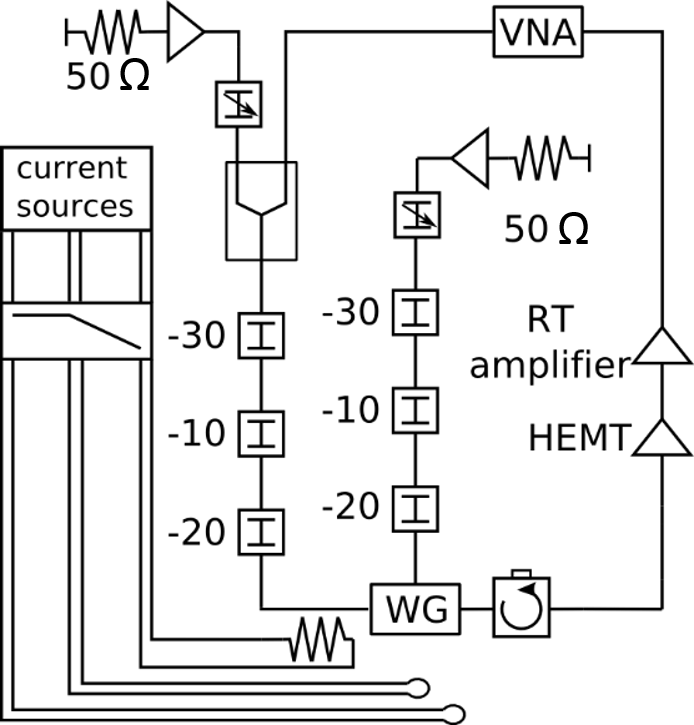}%
}\hfill
\subfloat[\label{fig:att_test_and_noise}]{%
\includegraphics[scale=0.35]{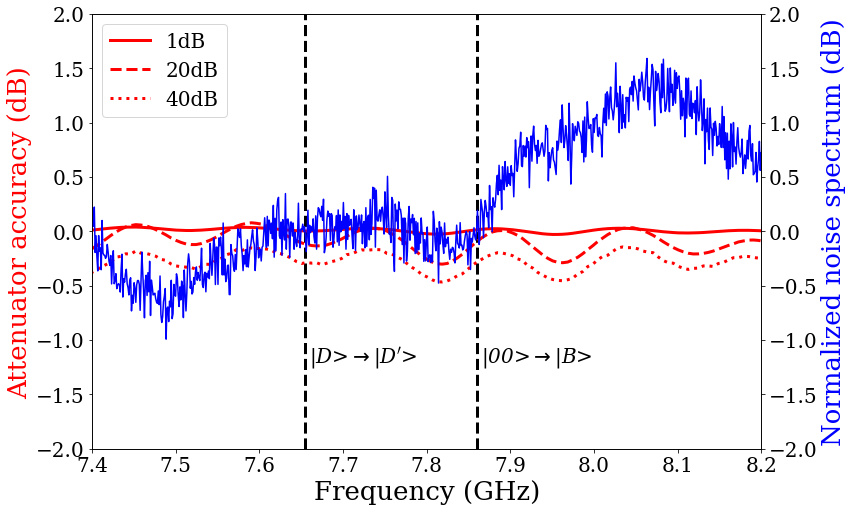}%
}\hfill
\caption{\label{fig:heater_pic}(a) Experimental setup. WG is the waveguide with the qubits inside. Noise sources (room temperature amplifiers terminated with 50~$\Omega$ at the input) are connected to the waveguide through variable attenuators. (b) Ripples of the variable attenuator and the normalized spectrum of the noise source in the frequency range of interest. The noise spectrum was normalized to its value at the $\ket{00}\rightarrow\ket{B}$ transition. Both $\ket{00}\rightarrow\ket{B}$ and $\ket{D}\rightarrow\ket{D^{\prime}}$ frequencies are shown with vertical dashed lines.}
\end{figure}

\section{Theoretical model}
\label{sec:theory_section}

We consider a model of two weakly anharmonic transmon qubits with frequency $\omega_1$ and $\omega_2$. The frequency of qubit $\text{Q}_2$ can be varied by applying a suitable flux bias. The qubits interact through a weak coupling capacitor and the qubit-qubit coupling energy is $g$. They are also weakly coupled to two different thermal environments, which we term ``global'' and ``local'', with temperatures respectively $T_\text{glob}$ and $T_\text{loc}$. The difference between the global and local baths lies in how they interact with the qubits: while the global bath is coupled with the same strength to each qubit (i.e., the coupling is symmetric), the local bath still interacts with both qubits but is more strongly coupled to one of them, say to qubit $\text{Q}_2$. Finally, the system is probed with a coherent drive tone with a frequency $\omega_\text{drive}$.

The free Hamiltonian of each transmon qubit as a weakly anharmonic oscillator is \cite{koch_charge-insensitive_2007,Blais2020}:
\begin{equation}
    \label{eqn:freeQubitHamil}
    H_{q_\alpha}=\hbar\omega_\alpha a_\alpha^\dagger a_\alpha +\hbar\beta_\alpha a_\alpha^\dagger a_\alpha^\dagger a_\alpha a_\alpha, \text{ with }\alpha=1,2.
\end{equation}
$\beta_\alpha$ is the anharmonicity of the qubit $\alpha$, and to take into account thermal effects we will consider multiple transmon levels. The free system Hamiltonian (including the probe) is:
\begin{equation}
    \label{eqn:freeSystemHam}
    H_S(t)=\sum_{\alpha=1,2}H_{q_\alpha}+H_\text{int}+H_\text{probe}(t),
\end{equation}
with
\begin{equation}
    \label{eqn:intDrive}
\begin{split}
    & H_\text{int}=g(a_1^\dagger a_2+a_1 a_2^\dagger),\qquad H_\text{probe}(t)=E_1 a_1^\dagger e^{-i\omega_\text{drive} t}+E_2 a_2^\dagger e^{-i\omega_\text{drive} t}+H.c.
\end{split}
\end{equation}
The free Hamiltonian of the thermal baths is given by:
\begin{equation}
    H_B=\sum_{j=\text{g},\text{l}} \sum_k \Omega_{\alpha,k} b_{j,k}^\dagger b_{j,k},
\end{equation}
where $j=\text{g},\text{l}$, labels the frequencies and operators of respectively the global and the local bath. Finally, assuming that the qubits and the baths are coupled through their quadratures, giving rise to dissipation \cite{cattaneo_engineering_2021}, the interaction Hamiltonian is given by:
\begin{equation}
    \label{eqn:HamIntQubitBath}
    H_{qb}=\mu \left[\sum_k \zeta_{\text{g},k} \sum_{\alpha=1,2}(a_\alpha^\dagger b_{\text{g},k}+a_\alpha b_{\text{g},k}^\dagger) + \sum_k \zeta_{\text{l},k}  \sum_{\alpha=1,2}  \lambda_{\alpha} (a_\alpha^\dagger b_{\text{l},k}+a_\alpha b_{\text{l},k}^\dagger)\right].
\end{equation}
 $\mu$ is the magnitude of the system-environment interaction. The weak coupling limit reads $\mu\ll\hbar\omega_\alpha$, i.e., the interaction energy is much lower than the qubit energies. Both $\zeta_{j,k}$ and $\lambda_\alpha$ are dimensionless parameters. The coefficients $\zeta_{j,k}$ define the spectral density of the baths $J(\omega)$ \cite{cattaneo_engineering_2021,cattaneo_bath-induced_2021}, which, for simplicity and according to experimental observations (see the considerations in Appendix~\ref{sec:spectrum_consideration}), we take as perfectly flat for both local and global baths. The global bath is (ideally) coupled to each qubit with the same weight. In contrast, the local bath is more strongly coupled to qubit $\text{Q}_2$ than to qubit $\text{Q}_1$. In the main text we have introduced the ratio between the local decay rates of the two qubits, which can be written as
 \begin{equation}
     k = \left(\frac{\lambda_2}{\lambda_1}\right)^2,
 \end{equation}
 with $\lambda_2>\lambda_1$.
The case with $k\rightarrow 1$ corresponds to the limit of a global symmetric bath, while $k\rightarrow +\infty$ is the limit of perfectly local bath, that is, the local environment is completely decoupled from qubit $\text{Q}_1$. 
 Finally,  the total Hamiltonian of system+environment is:
\begin{equation}
    \label{eqn:totHam}
    H=H_S(t)+H_B+H_{qb}.
\end{equation}
Note that the we can switch off the local and/or global bath by setting the coefficients respectively $\zeta_{l,k}$ and $\zeta_{g,k}$ to zero.

The dependence on time in Eq.~\eqref{eqn:totHam} can be eliminated through a suitable unitary transformation to write the Hamiltonian in a frame rotating with the driving frequency \cite{hofer_markovian_2017}. Then, we can derive a Markovian master equation for the state of the two qubits in the rotating frame. The Markovianity of the evolution is justified by the weak coupling between the qubits and the bath, and by the flat spectral density that leads to a vanishing ``memory time'' of the environment \cite{BreuerPetruccione}. Moreover, the master equation is \textit{local}, i.e., the jump operators are local operators acting on a single qubit only \cite{cattaneo_local_2019,Trushechkin2016}. This is justified by the fact that the qubit-qubit coupling energy and the driving energy are much lower than the system energy \cite{cattaneo_local_2019,hofer_markovian_2017,Gonzalez2017,Trushechkin2016}, $ g,E_1,E_2\ll \hbar\omega_1,\hbar\omega_2
$, and that the autocorrelation functions of the environment are a Dirac's delta in time due to the flat spectral density \cite{hofer_markovian_2017}. 

Transmon qubits are only weakly anharmonic, thus their higher levels can be easily populated as well at high enough temperature, making them qudits instead of proper qubits. For simplicity, let us consider the first $d$ levels of the transmon and truncate the rest, such that the eigenstates of the free transmon Hamiltonian are $\{\ket{j}_\alpha\}_{j=0}^{d-1}$ with energies $e_0^{(\alpha)},\ldots,e_{d-1}^{(\alpha)}$. The energies of each transmon level depend on the qubit frequency and on the anharmonicity according to Eq.~\eqref{eqn:freeQubitHamil}. Then, we can introduce the lowering operators:
\begin{equation}
    \sigma_{j,k}^{(\alpha)}=\sqrt{k}\ket{j}_\alpha\!\bra{k}, \text{ such that }    a_\alpha=\sum_{j=0}^{d-2} \sigma_{j,j+1}^{(\alpha)}.
\end{equation}
$\sigma_{j,j+1}^{(\alpha)}$ and $\sigma_{j+1,j}^{(\alpha)}$ are the jump operators \cite{BreuerPetruccione,cattaneo_local_2019} of the local master equation, respectively associated with the jump frequencies $\omega_j^{(\alpha)}$ and $-\omega_j^{(\alpha)}$. The jump frequencies are simply the difference in frequency between two adjacent energy levels, that is,
\begin{equation}
    \label{eqn:jumpFrequencies}
    \omega_j^{(\alpha)}=e_{j+1}^{(\alpha)}-e_{j}^{(\alpha)}.
\end{equation}
Finally, the master equation reads:
\begin{equation}
\label{eqn:masterEq}
    \frac{d}{dt}\rho_S(t)=-i[\tilde{H}_{q_\alpha}+H_{int}+H_{probe}(0),\rho_S(t)]+\mathcal{D}_\text{tot}[\rho_S(t)],
\end{equation}
with $\tilde{H}_{q_\alpha}=(\omega_\alpha-\omega_\text{drive})  a_\alpha^\dagger a_\alpha +\beta_\alpha a_\alpha^\dagger a_\alpha^\dagger a_\alpha a_\alpha$, and dissipator given by:
\begin{equation}
\label{eqn:dissipator}
\mathcal{D}_\text{tot}[\rho]=\sum_{\alpha,\alpha'=1,2}\sum_{j,k=0,\ldots,d-2}\left[\gamma_{\alpha j,\alpha' k}^\downarrow \mathcal{D}(\sigma_{j,j+1}^{(\alpha)},\sigma_{k,k+1}^{(\alpha')})[\rho]+\gamma_{\alpha j,\alpha' k}^\uparrow \mathcal{D}(\sigma_{j+1,j}^{(\alpha)},\sigma_{k+1,k}^{(\alpha')})[\rho]\right].
\end{equation}
We have introduced the notation $\mathcal{D}(a,b)[\rho]=a\rho b^\dagger-\frac{1}{2}\{b^\dagger a,\rho\}$ and the coefficients:
\begin{equation}
\label{eqn:decayRateFin}
   \gamma_{\alpha j,\alpha' k}^{\downarrow\uparrow}=\left[\Gamma^{\downarrow\uparrow}(\omega_j^{(\alpha)},T_\text{glob})+\Gamma^{\downarrow\uparrow}(\omega_{k}^{(\alpha')},T_\text{glob})\right]+ \lambda_\alpha\lambda_{\alpha'}\left[\Gamma^{\downarrow\uparrow}(\omega_j^{(\alpha)},T_\text{loc})+\Gamma^{\downarrow\uparrow}(\omega_{k}^{(\alpha')},T_\text{loc})\right],
\end{equation}
and we have employed the quantity
\begin{equation}
    \Gamma^{\downarrow\uparrow}(\omega,T)=\frac{\pi \gamma_0}{2}  \left(\coth(\frac{\omega}{2k_B T})\pm 1\right).
\end{equation}
The $+$ sign in the brackets corresponds to $\Gamma^\downarrow$ and analogously for $\Gamma^\uparrow$. $\gamma_0$ is a decay constant with the units of the inverse of time, which is proportional to the square of the coupling strenght $\mu^2$ and depends on the spectral density of the baths $J(\omega)$, which is flat. How to find $\gamma_0$ in capacitively coupled transmon qubits, for instance, we refer the reader to Refs.~\cite{cattaneo_bath-induced_2021,cattaneo_engineering_2021}. We note that the structure of the master equation in Eq.~\eqref{eqn:dissipator} is the same for both the global and local bath, while the expression for the contributions to the decay rates is different, as given by Eq.~\eqref{eqn:decayRateFin}. While probing the global bath only, we need to set $\lambda_\alpha=0$ to remove the local contribution.

We note that the master equation in Eq.~\eqref{eqn:masterEq} is in \textit{partial secular approximation} \cite{cattaneo_local_2019,Cattaneo2020,Jeske2015}, i.e., it couples jump operators associated with slightly different jump frequencies. This is true even if we tune the flux bias on qubit $\text{Q}_2$ such that its frequency matches that of qubit $\text{Q}_1$. Indeed, the anharmonicities $\beta_1$ and $\beta_2$ are never exactly the same, so the jump frequencies between higher transmon levels are slightly different for the two qubits. The master equation in partial secular approximation takes these small differences into account and is obtained from the Bloch-Redfield equation by removing the fast-rotating terms only, instead of removing all the non-resonant terms (``full secular'') as typically done in the standard derivation of the Lindblad master equation \cite{BreuerPetruccione}. The drawback of the partial secular approximation is that it may generate dynamics that are not exactly completely positive due to small negativity issues. However, recent works have shown that complete positivity is often recovered in the partial secular regime \cite{Cresser2017a,Farina2019} and, if small negativities appear, then they are not a real issue because they are of the order of the accuracy of the master equation \cite{cattaneo_local_2019}, see for instance Ref.~\cite{Hartmann2020}.

Finally, we employ the steady state of Eq.~\eqref{eqn:masterEq} to estimate the amplitude of the transmission measurements in the waveguide according to standard input-output theory \cite{Gardiner}. More specifically, we compute the transmission coefficient derived in Sec.~II B of Ref.~\cite{lalumiere_input-output_2013} for an ensemble of inhomogeneous atoms in waveguide QED. 

If only the symmetric global bath was present, then there would exist a dark subspace (or decoherence-free subspace \cite{Lidar2003}) corresponding to a subradiant transition between the qubits \cite{cattaneo_bath-induced_2021}. The subspace would not be populated in ideal and symmetric conditions. In contrast, the local bath breaks the symmetry generating the dark subspace and it can induce transitions thereto.

\newpage

\begin{figure*}[h!]
\subfloat[\label{fig:level_scheme_full}]{%
\includegraphics[scale=0.5]{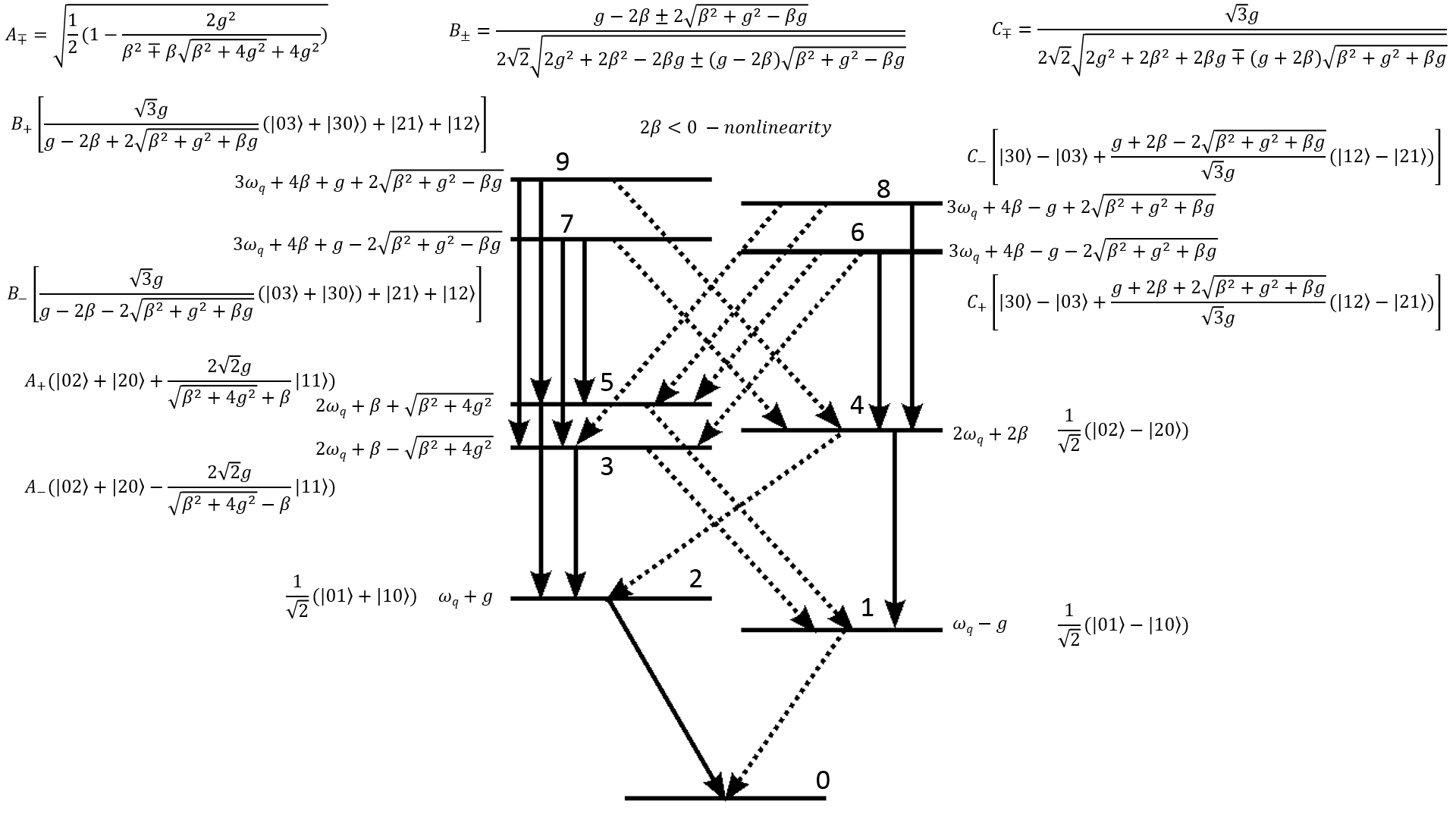}%
}\hfill
\subfloat[\label{fig:levels}]{%
\includegraphics[scale=0.5]{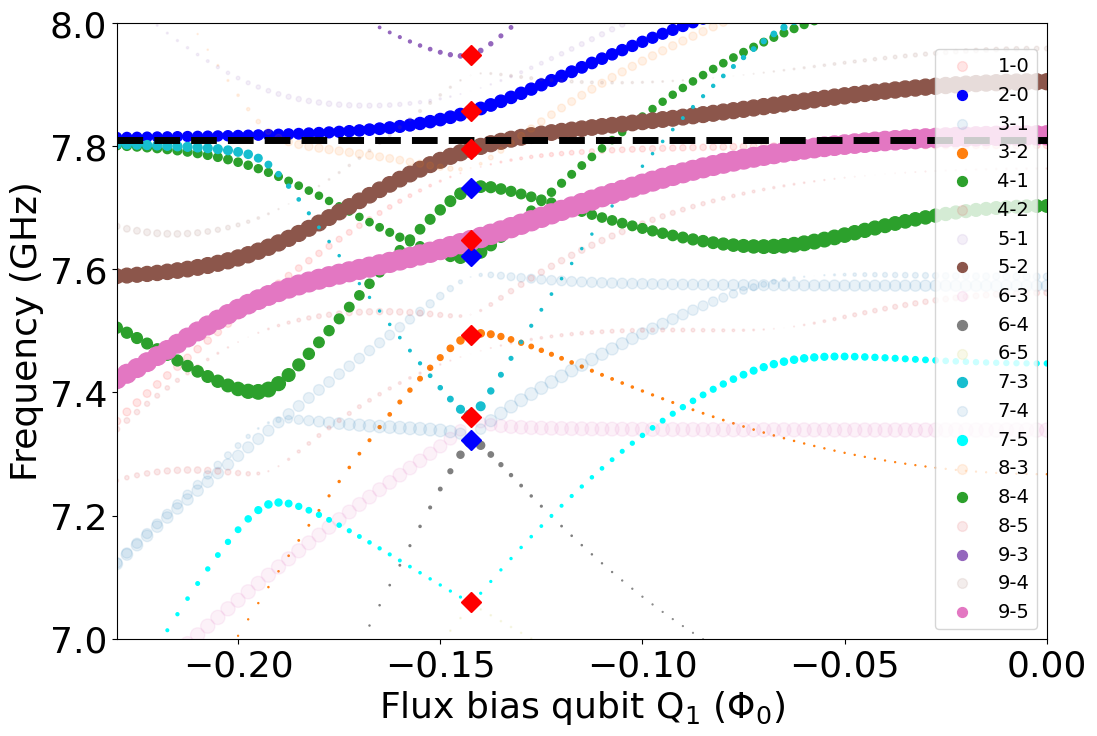}%
}\hfill
\caption{\label{fig:levels_and_lines} (a) Level scheme and analytical wave functions for 2 coupled transmons of the same frequency $\omega_q$ and anharmonicity $\beta<0$ with coupling energy $\hbar g$ up to 3 excitations. (b) Numerically calculated spectroscopy lines with varying qubit $\text{Q}_1$ frequency, while the frequency of qubit $\text{Q}_2$ is fixed at 7.81~GHz. The size of the dots is proportional to corresponding transitional dipole moment. Transparent lines correspond to the prohibited transitions when the qubit frequencies coincide. Diamond marks are the analytical values extracted from (a). Red diamonds correspond to transitions within the bright subspace, while blue diamonds are used for dark subspace transitions. Labels in the legend reflect mode numbers between which the transition occurs. In the resonant case these numbers coincide with levels numbering in (a).}
\end{figure*}

\section{Considerations on the spectral density}
\label{sec:spectrum_consideration}

Before describing our approach to the bath modelling we focus on the noise spectrum, as it is important to understand how the spectral density of the environment may affect the system dynamics.

In the main text we briefly discussed the energy levels of the coupled transmons up to 2 excitations, with a simplified scheme in Fig.~1(b). The same scheme up to 3 excitations is shown in Fig.~\ref{fig:levels_and_lines}\subref{fig:level_scheme_full}. Here we assume identical transmons of the same frequency $\omega_q$ and anharmonicity $\beta<0$. In the figure we specify the analytical energy values along with the corresponding wavefunctions. The normalization coefficients $A_{\pm}$, $B_{\pm}$ and $C_{\pm}$ are  also specified on the top of the figure. Allowed (with the waveguide fundamental mode) transitions are depicted with solid arrows while prohibited ones correspond to dashed arrows. The scheme shows a clear separation between bright (on the left) and dark (on the right) subspaces. In the main text we are using the 2-0 and 4-1 transitions as $\ket{00}\rightarrow\ket{B}$ and $\ket{D}\rightarrow\ket{D^{\prime}}$ respectively.

The frequencies of our transmons can be tuned independently and the transition lines seen in the experiment depend on the frequency detuning between the qubits. Fig.~\ref{fig:levels_and_lines}\subref{fig:levels} shows a plethora of lines one can expect to see in the experiment even when only 4 levels per transmon are taken into account. In this figure we model a scenario where qubit $\text{Q}_2$ has a stable frequency (dashed horizontal line) and qubit $\text{Q}_1$ frequency is changed by applying magnetic flux. In total the plot has 20 lines, 10 of them to prohibited transitions if the qubits are in resonance(transparent color, dashed arrows in the Fig.~\ref{fig:levels_and_lines}\subref{fig:level_scheme_full}) and another 10 allowed transitions (solid color, solid arrows in the Fig.~\ref{fig:levels_and_lines}\subref{fig:level_scheme_full}). Red and blue diamonds are the analytical values which were only calculated for the resonant case. Red diamonds correspond to the transitions within the bright subspace, while blue ones are for dark subspace transitions. For all transitions, marker sizes are proportional to the corresponding transitional dipole element. One can see that in the resonant case we expect all allowed transitions to be within 900~MHz between 7.05~GHz and 7.95~GHz. The highest and lowest frequency transitions (9-3 and 7-5) are only weakly allowed and correspond to population transfer between the 2-excitation and the 3-excitation manifolds. The lines of interest (2-0 and 4-1) connect the 1-excitation and 2-excitation manifolds and are separated only by the anharmonicity $\beta$=-230~MHz.
\begin{figure}[]
\centering
\includegraphics[scale=0.4]{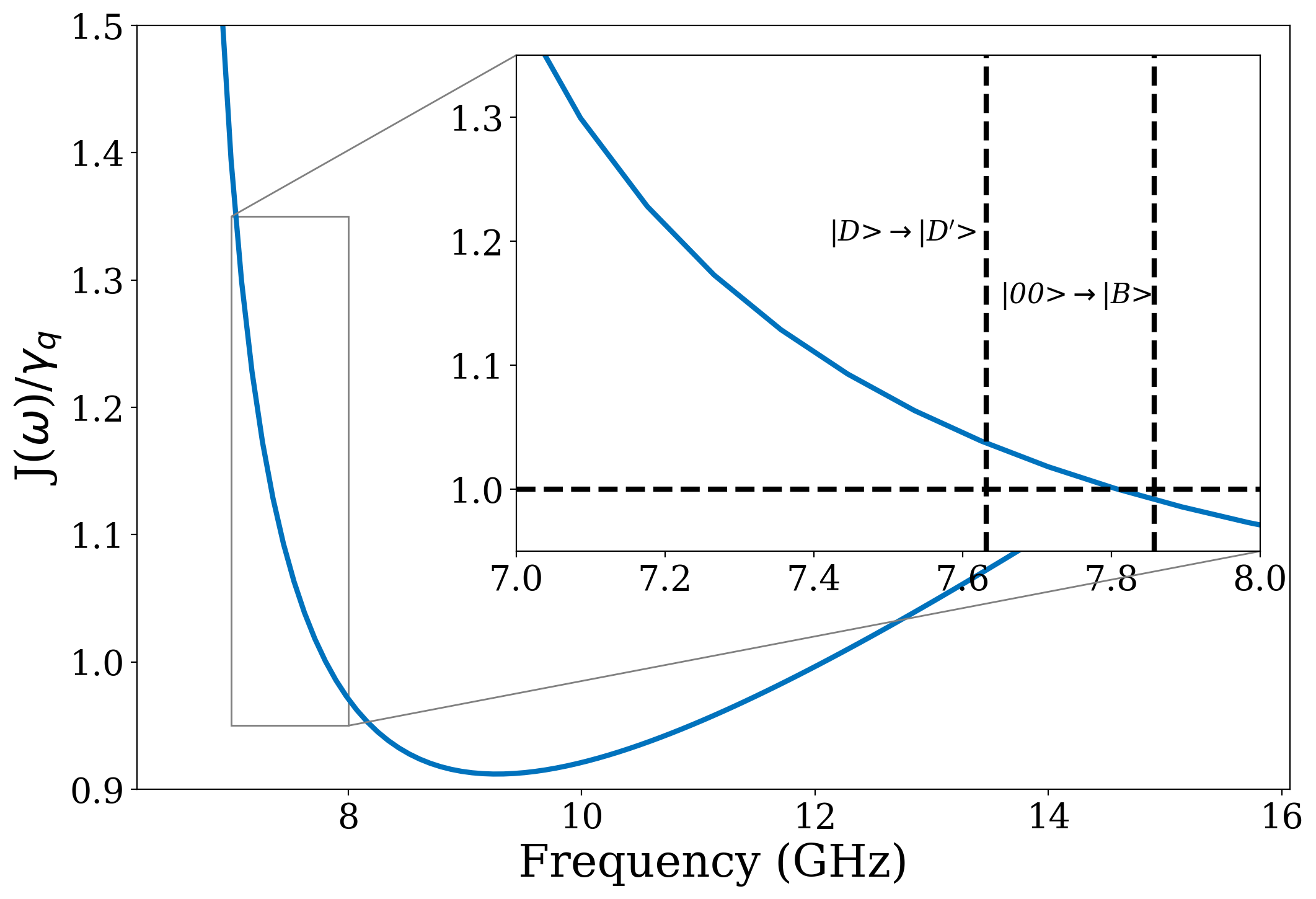}
\caption{\label{fig:J} Normalized spectral density of an infinite 3D rectangular cross section waveguide.}
\end{figure}
Having this in mind we can now consider whether taking into account a specific spectral density of the environment affects the results significantly. Fig.~\ref{fig:J} shows the normalized spectral density of modes inside the infinitely long waveguide with a cross section dimensions corresponding to our experimental case. One can see that the lines which we are using in the main text for the bright-dark diagrams can feel only a few \% variation in the spectral density of the environment. In general terms, the dynamics of the total system may depend on all the possible transition lines, and within our model the mismatch between the highest and lowest frequency lines due to the spectral density might be stronger (around 30\%). Nevertheless, we argue that these 2 lines only play a role in the dynamics at high temperatures where we do not expect precise result from our model due to not taken into account higher excitation manifolds. Other lines are closer to each other in frequency and therefore feel almost the same $J(\omega)$.

Another important consideration is that the line shown in Fig.~\ref{fig:J} diverges when the frequency approaches the cutoff frequency (around 6.55~GHz in our case) - this effect being relevant only for an infinitely long waveguide. In practice we are using commercial couplers connecting our waveguide with a standard 50~$\Omega$ coaxial lines. These couplers do not provide perfect impedance matching at the frequencies close to the cutoff and therefore the effective density of states seen by the qubits does not diverge at the cutoff and depends on details of the experimental setup. To take into account the effect of this mismatch (together with other imperfections in the transmission inevitably appearing in the input and output lines) on $J(\omega)$ seen by the qubits is a non-trivial task. This is aggravated by the fact that the couplers are non-identical and originally meant for room temperature usage. For an on-chip device utilizing a coplanar waveguide architecture this problem does not appear and one can assume an Ohmic spectral density whenever it is important. For this proof-of-principle device, in order to avoid additional assumptions and complications in the theoretical model, we decided to assume a flat spectral density (frequency independent), as it seems to describe the experimental results well enough. 

One of the difficulties of extracting temperature from transmission measurements is the necessity to remove the background of the transmission from the measurements. In experiments with 3D waveguides one has to use impedance matching elements at both ends of the waveguide, where the standard 50~$\Omega$ coaxial lines are connected to the waveguide. Because of imperfect impedance matching, the waveguide has low-Q resonant modes interacting with the strongly coupled qubits. As a result, the background one wishes to subtract depends on the qubit frequencies and populations. We attribute the mismatch between the best fit and the experimental data for the qubit $\text{Q}_1$ to this issue, which can be again be alleviated using a planar CPW architecture.

\section{Power calibration}
\label{sec:pow_cal}
In our theoretical model, described in Appendix~\ref{sec:theory_section}, we rely on the input-output theory \cite{lalumiere_input-output_2013} which requires to establish the relation between power and Rabi frequency. For our experimental setup we derive this relation from Autler-Townes experiments, i.e. by applying a coherent drive (through the waveguide or the side pin). The Hamiltonian of this drive can be written as 
\begin{equation}
\label{eqn:Rabi}
    H_{drive} = \frac{\Omega}{2} (\hat{a}^\dag + \hat{a}).
\end{equation}
Input-output theory yields the expression for the same Hamiltonian
\begin{subequations}
\begin{eqnarray}
H_{drive} =&E \hat{a}^\dag + E^\ast \hat{a},\label{AT1}
\\
E(t) =& -\textnormal{i} \sqrt{\frac{\gamma_{qubit}\cdot\omega_\text{drive}}{2\omega_{qubit}}}  D e^{-\textnormal{i} \omega_\text{drive} t}\label{AT2}
\\
D=& \textnormal{calibration} \sqrt{10^{\textnormal{drive[dBm]}/10}}\label{AT3}
\end{eqnarray}
\end{subequations}

The driving strength, $D$, is the fit parameter when treating the power sweep. In resonance ($\omega_\text{drive} = \omega_{qubit}$) we obtain the following relation between the fit parameter $D$ and the Rabi frequency $\Omega$ defined by Eq.~\eqref{eqn:Rabi}
\begin{equation}
    \Omega = 2 \sqrt{\gamma_{qubit}} D.
\end{equation}
Extracting the calibration for 3 different cases (global bath, side pin qubit $\text{Q}_1$ and side pin qubit $\text{Q}_2$) is shown in Fig.~\ref{fig:AT}. 

\begin{figure}[t]
\centering
\subfloat[\label{fig:AT_global_A}]{%
\includegraphics[scale=0.28]{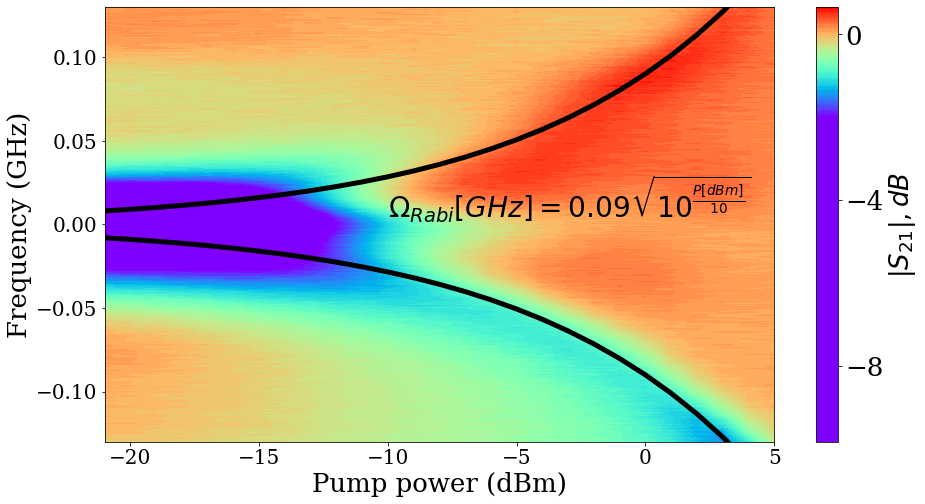}%
}
\subfloat[\label{fig:AT_pin_A}]{%
\includegraphics[scale=0.28]{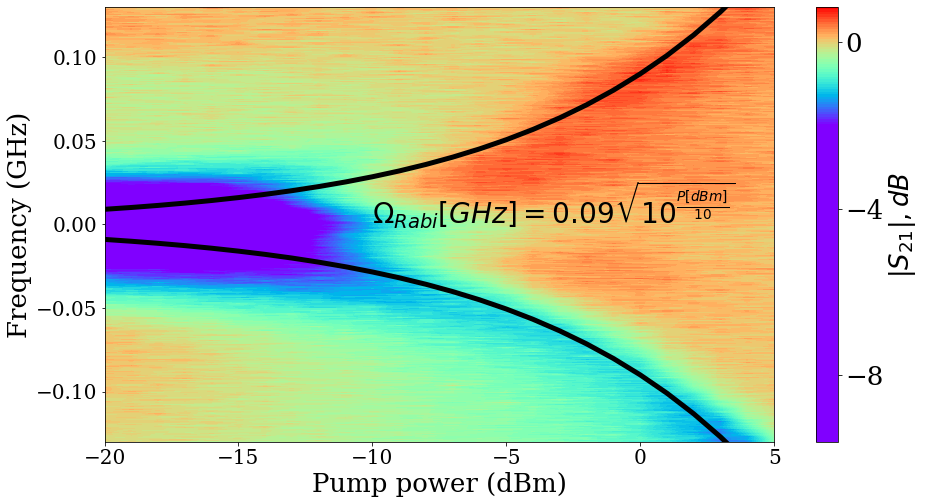}%
}\\
\subfloat[\label{fig:AT_pin_B}]{%
\includegraphics[scale=0.28]{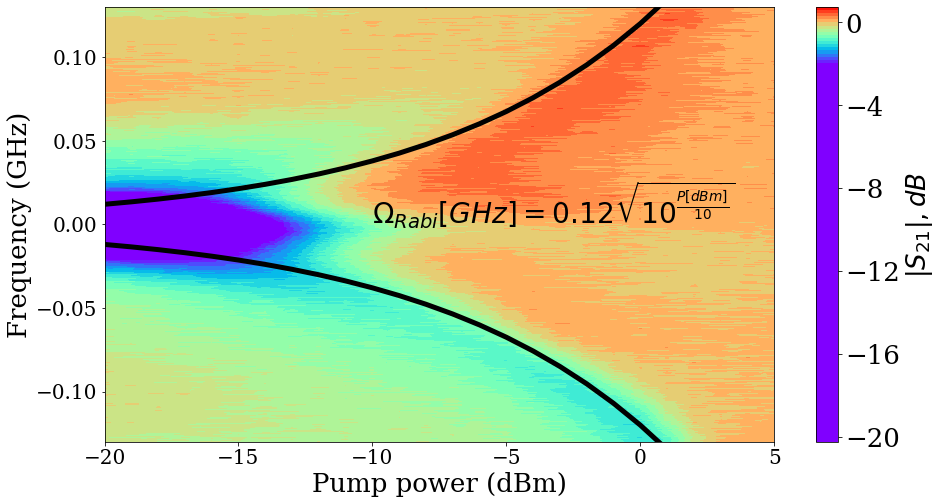}%
}\hfill
\caption{\label{fig:AT} Power calibration utlizing Autler-Townes splitting with the signal applied (a) through the waveguide mode (coupled symmetrically to both transmons), (b) through the side pin for the transmon $\text{Q}_1$, and (c) through the side pin for the transmon $\text{Q}_2$. The colour corresponds to transmission through the waveguide measured with a weak probe signal.}
\end{figure}
\mbox{}
\clearpage

\section{Fitting single qubit experiments with local bath}
\label{sec:local_fit_section}
We discuss now our technique for extracting independently side pin coupling and local bath temperature. The Autler-Towns experiment alone does not allow for the extraction of the coupling coefficient of a qubit to the drive line. For the global bath the situation is simpler as the probe tone couples with the same coupling constant as the drive. 

In principle, the calibration coefficient from the equation \eqref{AT3} can be extracted separately from $\sqrt{\gamma_{qubit}}$, but in practice it requires a rather rigorous calibration of all impedance mismatches in input and output lines, including mismatches on the input and output of the waveguide. These mismatches can be frequency and temperature dependent and calibrating them with the necessary precision is a possible but tedious task. Instead in this proof-of-principle experiment we chose a different strategy. From the Autler-Towns experiments with the side pin for qubits $\text{Q}_1$ and $\text{Q}_2$ we can extract the ratio $\sqrt{\gamma_{loc1}/\gamma_{loc2}}$. Using the side pin as a source for a local bath we follow both $\ket{g}\rightarrow\ket{e}$ and $\ket{e}\rightarrow\ket{f}$ transitions (see Fig.~\ref{fig:side_pin_single_qubit}). The curves for the $ket{g}\rightarrow\ket{e}$ transitions were fitted jointly for both qubits. 
Two fitting parameters were used in the procedure: $\alpha_{loc}$ and $\gamma_{loc1}$. Cable attenuation and impedance mismatches between cables are included in $\alpha_{loc}$, which was only allowed to vary around $\alpha_{glob}$, which in turn we already extracted during the global bath fitting. The second fitting parameter $\gamma_{loc1}$, which includes the large impedance mismatch between the coax cable with the pin and the waveguide, was allowed to vary within 3 orders of magnitude around the value expected from the device geometry. The local bath temperature values $T_{loc}=T_{res}+\alpha_{loc}\cdot P$ obtained from the fit correspond to what we already expect from the setup cabling and the considerations above. Results for the$\ket{e}\rightarrow\ket{f}$ transitions were not used in the fit and fitting curves in Fig.~\ref{fig:side_pin_single_qubit}\subref{fig:local_2q_fit_ef} just show how they appear in the model with all parameters fixed after the $\ket{g}\rightarrow\ket{e}$ fit. The values of $\alpha_{loc}$ and $\gamma_{loc1}$ obtained through this procedure were used to produce the 2 qubit local bath theory lines for Figs.~2(c) and 2(d) of the main text.      
\begin{figure}[]
\subfloat[\label{fig:local_2q_fit}]{%
\includegraphics[scale=0.32]{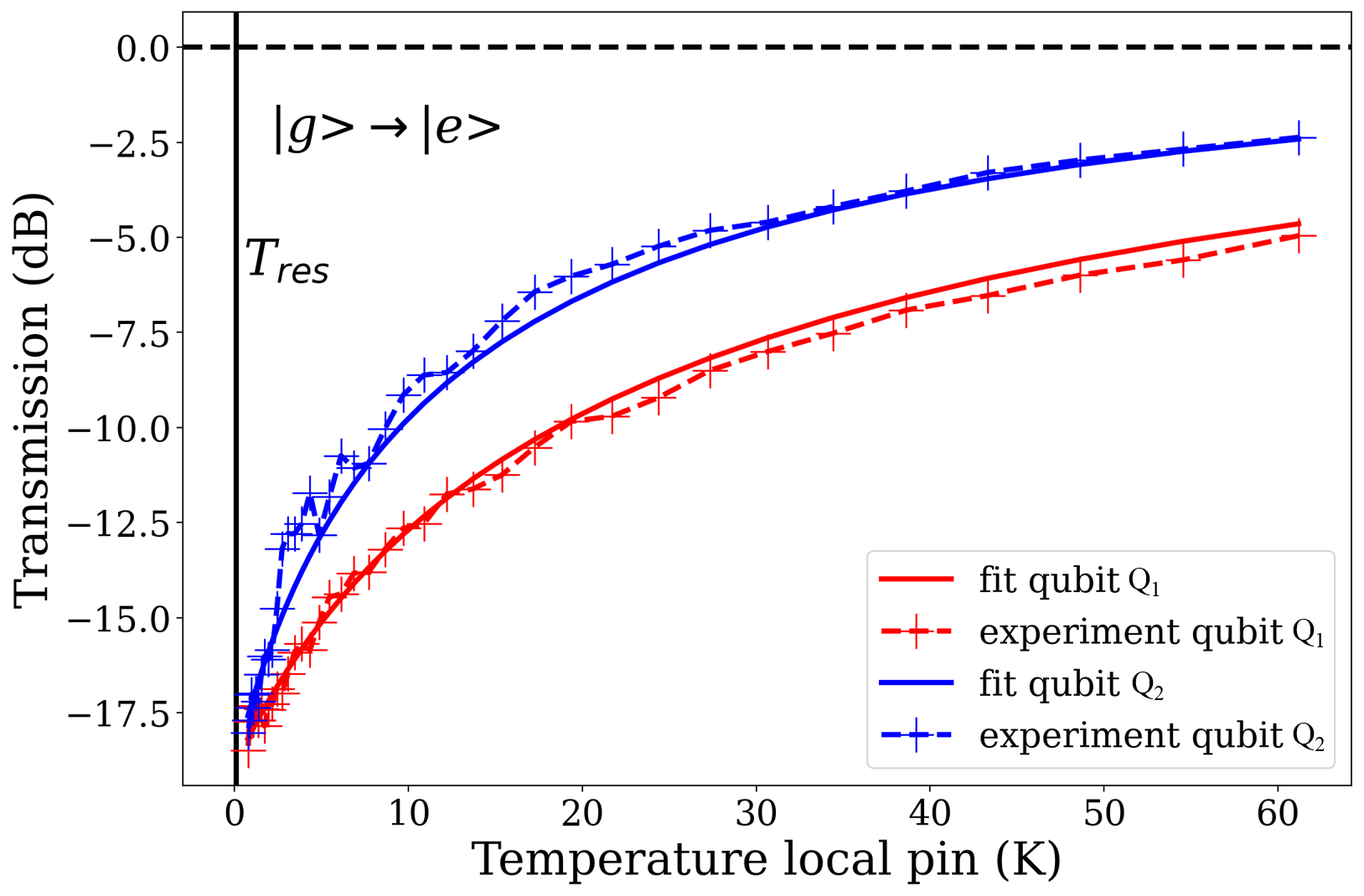}%
}
\subfloat[\label{fig:local_2q_fit_ef}]{%
\includegraphics[scale=0.32]{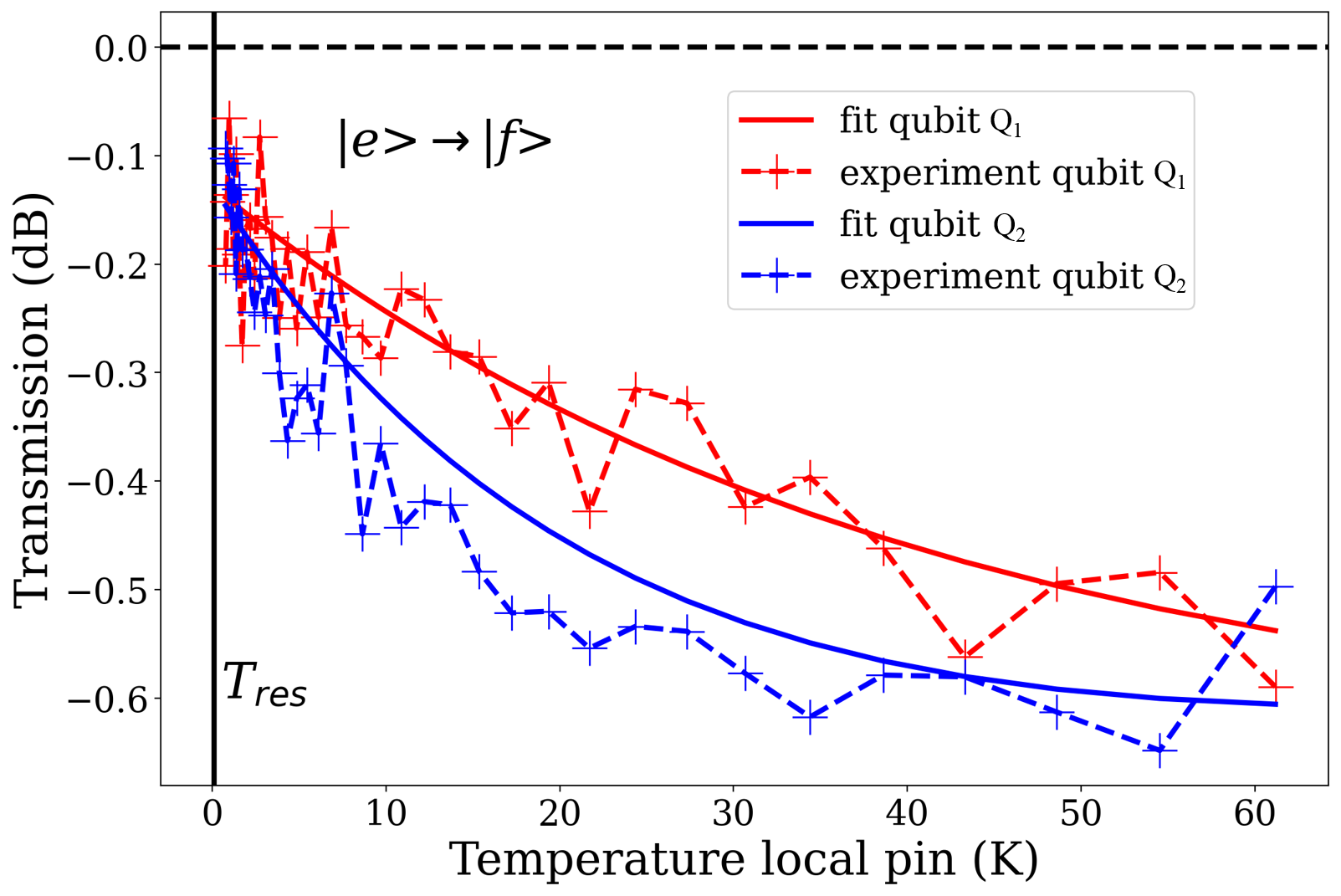}%
}\hfill
\caption{\label{fig:side_pin_single_qubit} Result of a single qubit experiments with a side pin as a local bath fitted together (see the text): (a) $\ket{g}\rightarrow\ket{e}$ transition and (b) $\ket{e}\rightarrow\ket{f}$.}
\end{figure}

\section{Further theoretical results}
\label{sec:model_results}
In this section we briefly discuss some relevant results one can obtain with the theoretical model, which were not included in the main text.
\begin{figure}[h]

\includegraphics[scale=0.3]{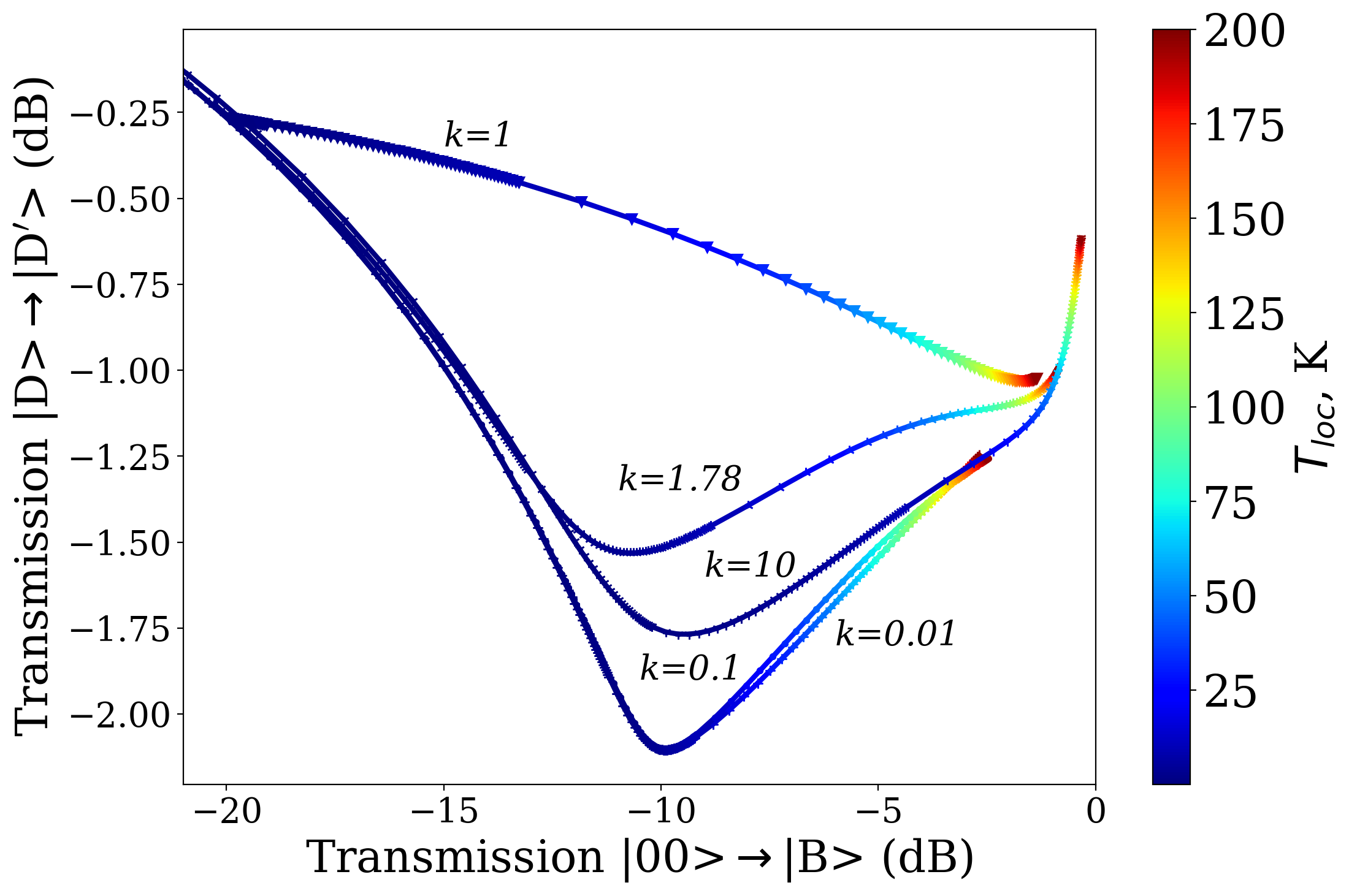}
\caption{\label{fig:k_cuts} Modelled bright-dark diagram for different coupling asymmetry of the side pin. The curves corresponding to the side pin coupled only to one of the qubits ($k\ll1$ or $k\gg1$) are distinguishable from the perfectly global case $k=1$. Points correspond to $T_{loc}$ changing from 0 to 200~K.} 
\end{figure}
One question appearing from Fig.~2(d) of the main text is how having a purely local (coupled only to one of the qubits) bath would change our bright-dark diagrams. Fig. \ref{fig:k_cuts} shows these diagrams for different balance parameter $k=\gamma_{loc2}/\gamma_{loc1}$. One can see that the $\ket{D}\rightarrow\ket{D^{\prime}}$ transmission dip is the shallowest when $k=1$, i.e. in a purely global case, while it is the deepest when $k\rightarrow0$ or $\infty$. Anyway, the dark-bright diagram for the latter scenario does not differ  dramatically in comparison with our experimental situation $k=1.78$. The reason why the $k=1$ line is not perfectly flat is that there are allowed transitions between higher levels corresponding to frequencies close to $\ket{D}\rightarrow\ket{D^{\prime}}$ (see line 9-5 from Fig.
\ref{fig:level_scheme_full}).
\begin{figure}[t]
\centering
\subfloat[\label{fig:fq_cuts}]{%
\includegraphics[scale=0.32]{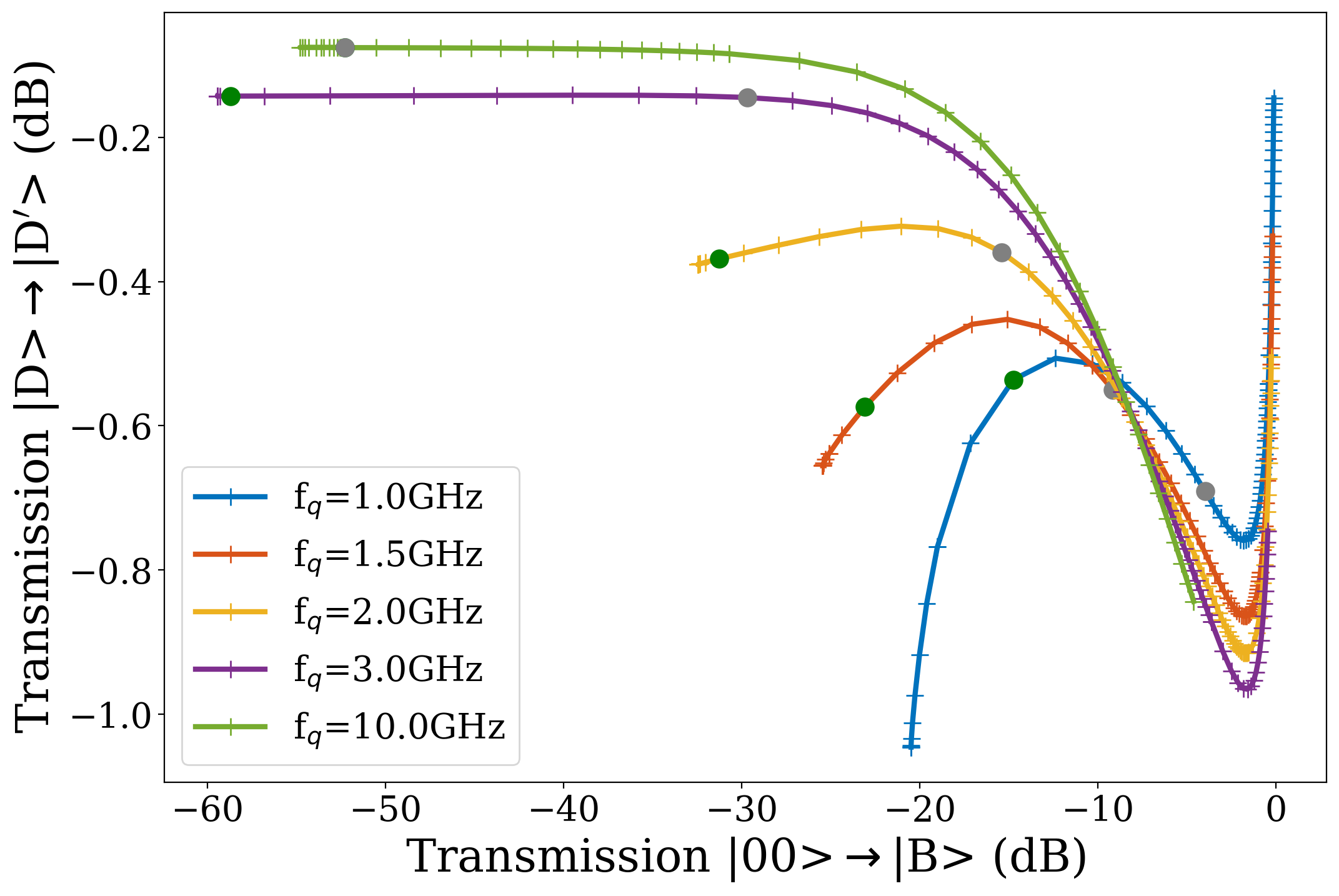}%
}
\subfloat[\label{fig:fq_cuts_no_T_loc}]{%
\includegraphics[scale=0.32]{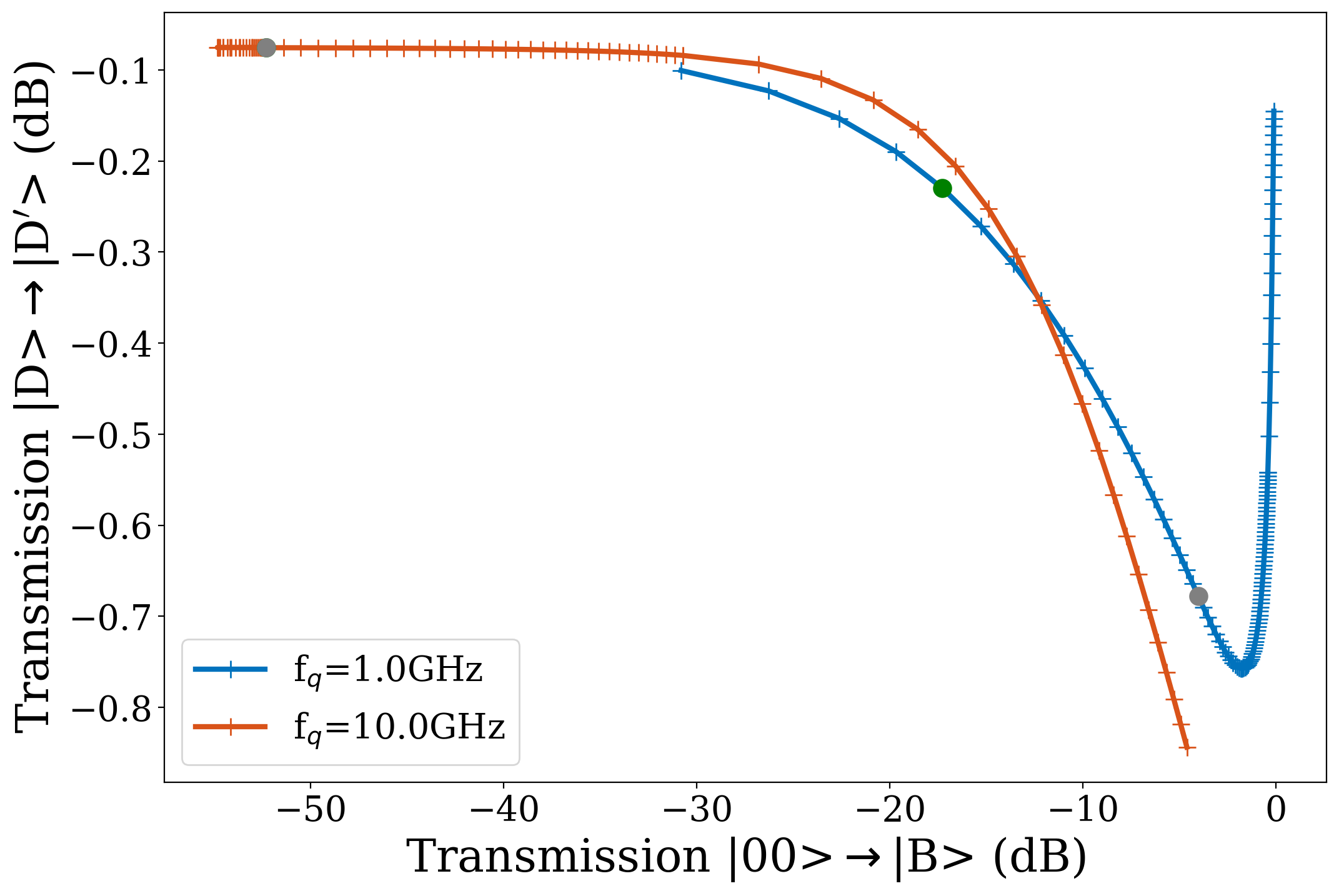}%
}

\caption{\label{fig:model_res} Result of single qubit experiments with a side pin as a local bath fitted jointly(see the text): (a) $\ket{g}\rightarrow\ket{e}$ transition and (b) $\ket{e}\rightarrow\ket{f}$.}
\end{figure}

Fig. \ref{fig:model_res}(a) is an addition to Fig.~3 in the main text and shows how varying the transmon frequency changes the bright-dark diagram. As before, the global bath temperature is being swept from 1 to 300~mK and $T_{loc}$ is fixed at 30~mK, while the other parameters correspond to the experimental case discussed in the main text, i.e. $\gamma_{loc1}=\gamma_{loc2}$ and $\gamma_{loc1}/2\pi=1$~kHz. Green circles on the lines correspond to $T_{glob}=14$~mK and gray circles are at $T_{glob}=30$~mK. One can see that lower frequency qubits react to the presence of the local bath even if it is weakly coupled and relatively cold. 

Fig.~\ref{fig:model_res}\subref{fig:fq_cuts_no_T_loc} corresponds to the same situation but the local bath temperature is fixed at a very low value of 1~mK. We depict here only 2 lines in comparison with Fig.~\ref{fig:model_res}\subref{fig:fq_cuts}, as they mostly overlap. In this case the local bath effectively acts as a weakly coupled relaxation channel, and regardless of the transmons frequency the bright-dark diagram looks qualitatively the same.  

We expect that measuring these lines in an experiment we should allow to shed some light on the source of the residual excitation in a modern transmon systems.

\end{widetext}
\end{document}